\documentclass{ptephy_my}

\usepackage{amsmath}
\usepackage{amsthm}
\usepackage{amssymb}

\usepackage{graphicx}
\usepackage{xcolor}

\usepackage{url}
\usepackage{bm}
\usepackage{natbib}

\graphicspath{{./Figures_reduced/}}

\numberwithin{equation}{section}

\newcommand{\bC}{\ensuremath{\mathbb{C}}}

\newcommand{\bH}{\ensuremath{\mathbb{H}}}

\newcommand{\bR}{\ensuremath{\mathbb{R}}}

\newcommand{\bZ}{\ensuremath{\mathbb{Z}}}

\newcommand{\scA}{\ensuremath{\mathcal{A}}}
\newcommand{\scB}{\ensuremath{\mathcal{B}}}

\newcommand{\scH}{\ensuremath{\mathcal{H}}}

\newcommand{\scN}{\ensuremath{\mathcal{N}}}
\newcommand{\scO}{\ensuremath{\mathcal{O}}}

\newcommand{\scQ}{\ensuremath{\mathcal{Q}}}

\newcommand{\scT}{\ensuremath{\mathcal{T}}}

\newcommand{\scW}{\ensuremath{\mathcal{W}}}

\newcommand{\bea}{\begin{equation}\begin{aligned}}
\newcommand{\eea}{\end{aligned}\end{equation}}
\newcommand{\beq}{\begin{eqnarray}}
\newcommand{\eeq}{\end{eqnarray}}

\newcommand{\sfx}{\ensuremath{\mathsf{y}}}

\newcommand{\sfu}{\ensuremath{\mathsf{u}}}
\newcommand{\sfp}{\ensuremath{\mathsf{p}}}

\newcommand{\Teichmuller}{Teichm\"{u}ller }
\newcommand{\Li}{\ensuremath{\mathrm{Li}}}

\newcommand{\tsigma}{\ensuremath{\tilde{\sigma}}}

	\def\P{{\Bbb{P}}}

\def\fpp#1#2{{\frac{\partial{#1}}{\partial{#2}} }}

\def\S{{\mathcal{S}}}

\def\P{{\Pi}}

\def\>{{\geq }}
\def\<{{\leq }}

\def\half{\frac{1}{2}}

\def\i{\sqrt{-1}}

\begin{document}

\title{$\scN=2$ Theories from Cluster Algebras}
\author{\name{\fname{Yuji} \surname{Terashima}}{1}, and \name{\fname{Masahito}
\surname{Yamazaki}}{2}}

%%%\name{}{} Insert author name in first group and
%%% affiliation number in second group
%%% \fname for author firstname
%%% \surname for author surname
%%% \midname for author middle names

\address{\affil{1}{Department of Mathematics, Tokyo Institute of Technology, Tokyo
 152-8551, Japan
}
\affil{2}{Princeton Center for Theoretical Science, Princeton University, NJ 08544, USA}
%\email{xxxx@xxxx.ac.jp}
}
%%%\affil{}{} Insert affiliation number in first group and
%%% address in second group
%\address{First author address, second author address, and
%third author address \email{xxxx@xxxx.ac.jp}}

\begin{abstract}

We propose a new description of 3d $\scN=2$ theories
which do not admit conventional Lagrangians.
Given a quiver $Q$
and a mutation sequence $\bm{m}$ on it,
we define a 3d
$\scN=2$ theory $\scT[(Q,\bm{m})]$
in such a way that the $S^3_b$ partition function of the theory
coincides with the \emph{cluster partition function} 
defined from the pair $(Q, \bm{m})$.

Our formalism includes the case where 
3d $\scN=2$ theories arise
from the compactification of the 6d $(2,0)$ $A_{N-1}$
theory on a
large class of 3-manifolds $M$, 
including complements of arbitrary links in $S^3$.
In this case the quiver is defined from a
2d ideal triangulation, the mutation sequence represents an element of the
mapping class group, and the 3-manifold is equipped with a canonical
ideal triangulation.
Our partition function then
coincides with that of the holomorphic part of the $SL(N)$ Chern--Simons partition function on $M$.

\end{abstract}
\subjectindex{B10, B16, B34} %%% for Subject index
\maketitle

%%%%%%%%%%%%%%%%%%%%%%%%%%%%%%%%%%%%%%%%%%%%%
\section{Introduction}\label{sec.intro}
%%%%%%%%%%%%%%%%%%%%%%%%%%%%%%%%%%%%%%%%%%%%%

It has recently been discovered
\cite{Terashima:2011qi,Dimofte:2011ju,Cecotti:2011iy} 
(see also earlier works
\cite{Drukker:2010jp,Dimofte:2010tz,Hosomichi:2010vh})
that 
there exists a beautiful correspondence (``3d/3d correspondence'')
between the 
physics of 3d $\scN=2$ gauge theories
and the geometry of 3-manifolds.
The latter, in more physical language, is the study of 
analytic continuation of 3d pure Chern--Simons $SU(2)$ gauge theory 
into a non-compact gauge group $SL(2)$ \cite{Witten:1989ip,Witten:2010cx}. More
quantitatively, one of the consequences of this correspondence
is that given a 3-manifold $M$
there is a corresponding 3d $\scN=2$ theory $\scT[M]$
such that the partition functions of the two theories coincide \cite{Terashima:2011qi}:
\begin{align}
Z^{\textrm{3d } \scN=2}_{\scT[M]}[S^3_b]=Z^{\textrm{Chern--Simons}}[M] \ ,
\label{ZZ}
\end{align}
where the left-hand side is the 3-sphere 
partition function \cite{Kapustin:2009kz,Jafferis:2010un,Hama:2010av}
with a 1-parameter deformation by $b$ \cite{Hama:2011ea}, and 
the right-hand side is the holomorphic part of the $SL(2)$ Chern--Simons
theory with level $t$. The two parameters $b$ and $t$ are related by 
$b^2\sim 1/t$.
\footnote{
Note that in correspondence \eqref{ZZ} 
the same data appears in rather different guises on the two sides.
For example, the geometry of $M$ for the right hand side 
determines the choice of the theory $\scT[M]$ itself on the left. Similarly,
the deformation of the geometry, the parameter $b$, on the left hand side
is translated into a parameter of the Lagrangian on the right.
}
\footnote{The first evidence for this conjecture \cite{Terashima:2011qi} came from a chain of arguments
involving quantum Liouville and \Teichmuller theories.
The semiclassical ($t\to \infty$) expansion of the 
right-hand side of Eq.\ \eqref{ZZ} reproduce
hyperbolic volumes and Reidemeister torsions of 3-manifolds
\cite{Terashima:2011xe,Nagao:2011aa}. See 
\cite{Dimofte:2011jd,Dimofte:2011py,Teschner:2012em,Gang:2012ff,Cordova:2012xk}
for further developments in the 3d/3d correspondence.}

The correspondence \eqref{ZZ} provides a fresh perspective on 
the systematic study of a large class
of 3d $\scN=2$ theories, and relations between them.
An arbitrary hyperbolic 3-manifold could be constructed by 
gluing ideal tetrahedra, and correspondingly we could construct complicated 
3d $\scN=2$ theories starting from free $\scN=2$ chiral multiplets.
Gluing ideal tetrahedra is translated into the gauging of the global
symmetries,
and the change of polarization into an $Sp(2N, \bZ)$ action
\cite{Kapustin:1999ha,Witten:2003ya}
on 3d $\scN=2$ theories.
The 2-3 Pachner move, which represents the change of 
the ideal triangulation of the 3-manifold, is translated into the
3d $\scN=2$ mirror symmetry \cite{Dimofte:2011ju}.   

Despite the beauty of this correspondence, 
we should keep in mind limitations of the correspondence
\eqref{ZZ} --- not all 3d $\scN=2$ theories are of the form $\scT[M]$.
The natural question is whether we can generalize the correspondence to 
a larger class of 3d $\scN=2$ theories beyond those associated with
3-manifolds, or more generally 
whether there are any geometric structures in the 
``landscape'' or ``theory space'' of 3d $\scN=2$ theories.
To answer these questions it is crucial to 
extract the essential ingredients from the correspondence \eqref{ZZ}.

Our answer to this question is that it is the mathematical structures of
\emph{quiver mutations} and \emph{cluster algebras} which are essential for the correspondence \eqref{ZZ}.
We study 3d $\scN=2$ gauge theories $\scT[M]$ for a large class of
hyperbolic 3-manifolds, including arbitrary link complements in $S^3$.\footnote{
Suppose that we have a 3-manifold $M$, and a link $L$ inside.
The complement of a link is defined as a complement of a thickened link.
More formally, the link
complement is a complement of the tubular neighborhood
$N(L)$ of $M$, i.e, $M\backslash N(L)$.
By construction the boundary of the link complement is 
a disjoint union of 2d tori.
}
The present paper
generalizes the previous works on
this subject by the authors \cite{Terashima:2011qi,Terashima:2011xe},
and is a companion to the previous paper with K.~Nagao \cite{Nagao:2011aa}.
It is also closely related to \cite{Cecotti:2011iy} (see also \cite{Cecotti:2010fi,Cordova:2012xk}).\footnote{
However, it is worthing emphasizing that their braid (branched locus)
is \emph{not} our braid; see further comments in section \ref{sec.hyperbolic}.
}

\begin{figure}[htbp]
\centering{\includegraphics[scale=0.23]{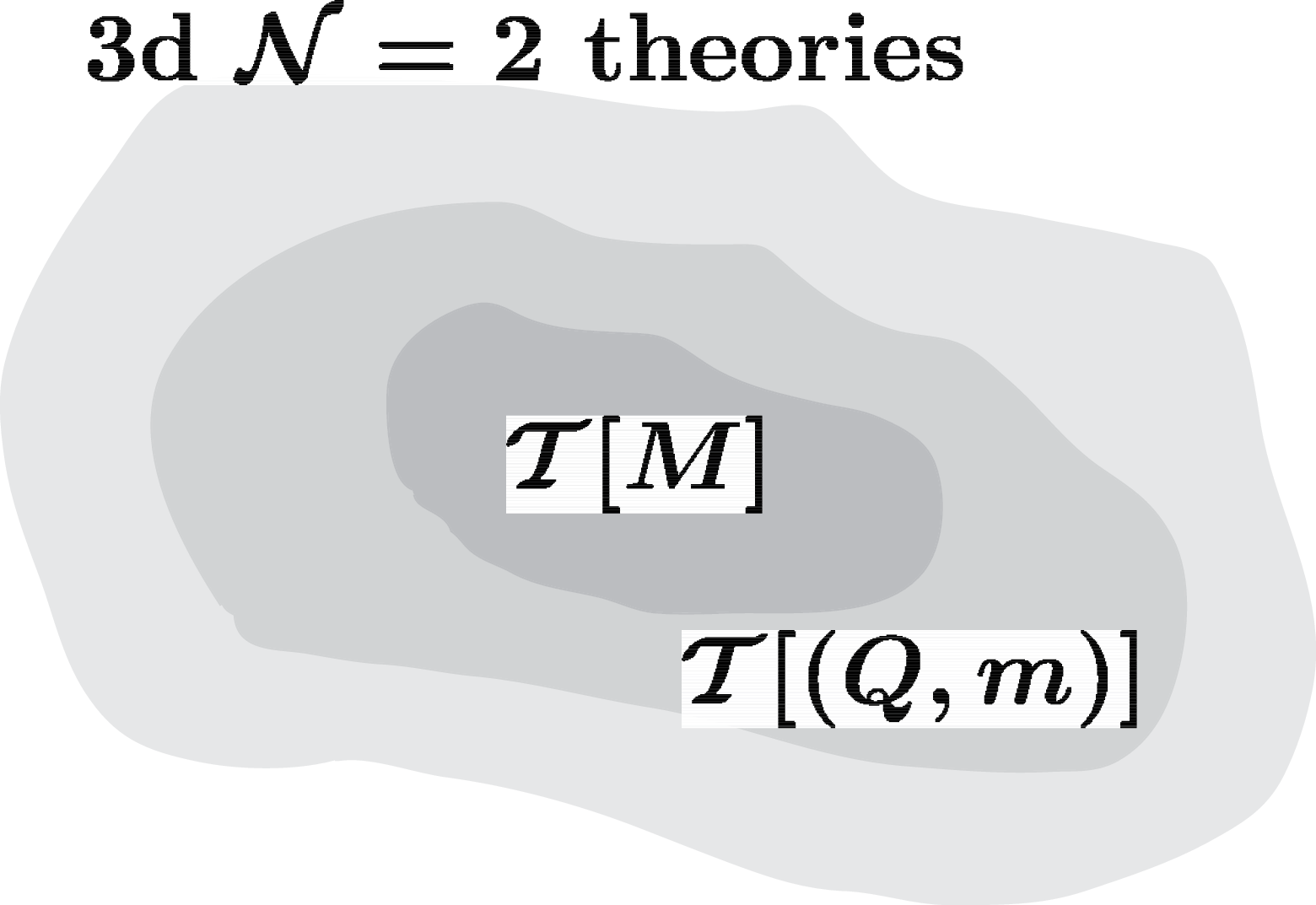}}
\caption{The class of 3d $\scN=2$
theories (denoted by $\scT[(Q,\bm{m})]$)
is a special subset of general 3d $\scN=2$ theories,
and is large enough to contain theories $\scT[M]$
dual to the geometry 3-manifolds.
Little is known at present about the detailed properties of the theories
$\scT[(Q,\bm{m})]$,
when the theory cannot be written in the form $\scT[M]$.
}
\label{fig.landscape}
\end{figure}

Our approach turns out to be much more general than \eqref{ZZ},
and includes
theories associated with $SL(N)$ Chern--Simons theories on
3-manifolds (cf.\ \cite{FockGoncharovHigher}), or more general theories not associated with 3-manifolds 
(Figure \ref{fig.landscape}).\footnote{
All our theories are contained in the theories of ``class $\mathcal{R}$'' in
\cite{Dimofte:2011py}. 
For comparison one might be tempted to call the
theories $\scT[M]$ to be of ``class $\mathcal{M}$'' ($\mathcal{M}$ for manifold)
and theories $\scT[(Q,\bm{m})]$ to be of ``class $\mathcal{C}$'' ($\mathcal{C}$ for cluster algebras).
}
We define a 3d $\scN=2$ theory $\scT[(Q, \bm{m})]$
for a pair of a quiver $Q$ and a mutation sequence $\bm{m}$ on
it,\footnote{The theory in addition depends on the choice of the
boundary condition, as will be explained in section \ref{sec.identification}.} 
satisfying the relation
\begin{align}
Z^{\textrm{3d } \scN=2}_{\scT[(Q, \bm{m})]}[S^3_b]=Z^{\rm cluster}_{(Q, \bm{m})} \ ,
\label{ZTQ}
\end{align}
where the right-hand side is the \emph{cluster partition function}
defined in this paper. The right-hand side contains a quantum parameter $q$, which is 
related to the parameter $b$ on the left-hand side by the relation \eqref{qb}.
We can think of the pair $(Q, \bm{m})$ as the defining data specifying the matter contents 
of 3d $\scN=2$ theories,
which in general do not have Lagrangian descriptions.
It will be an exciting possibility to explore the properties of these
theories further.

%%%%%%%%%%%%%%%%%%%%%%%%%%%%%%%
\subsection*{Summary}
%%%%%%%%%%%%%%%%%%%%%%%%%%%%%%

\begin{figure}[htbp]
\centering\includegraphics[scale=0.3]{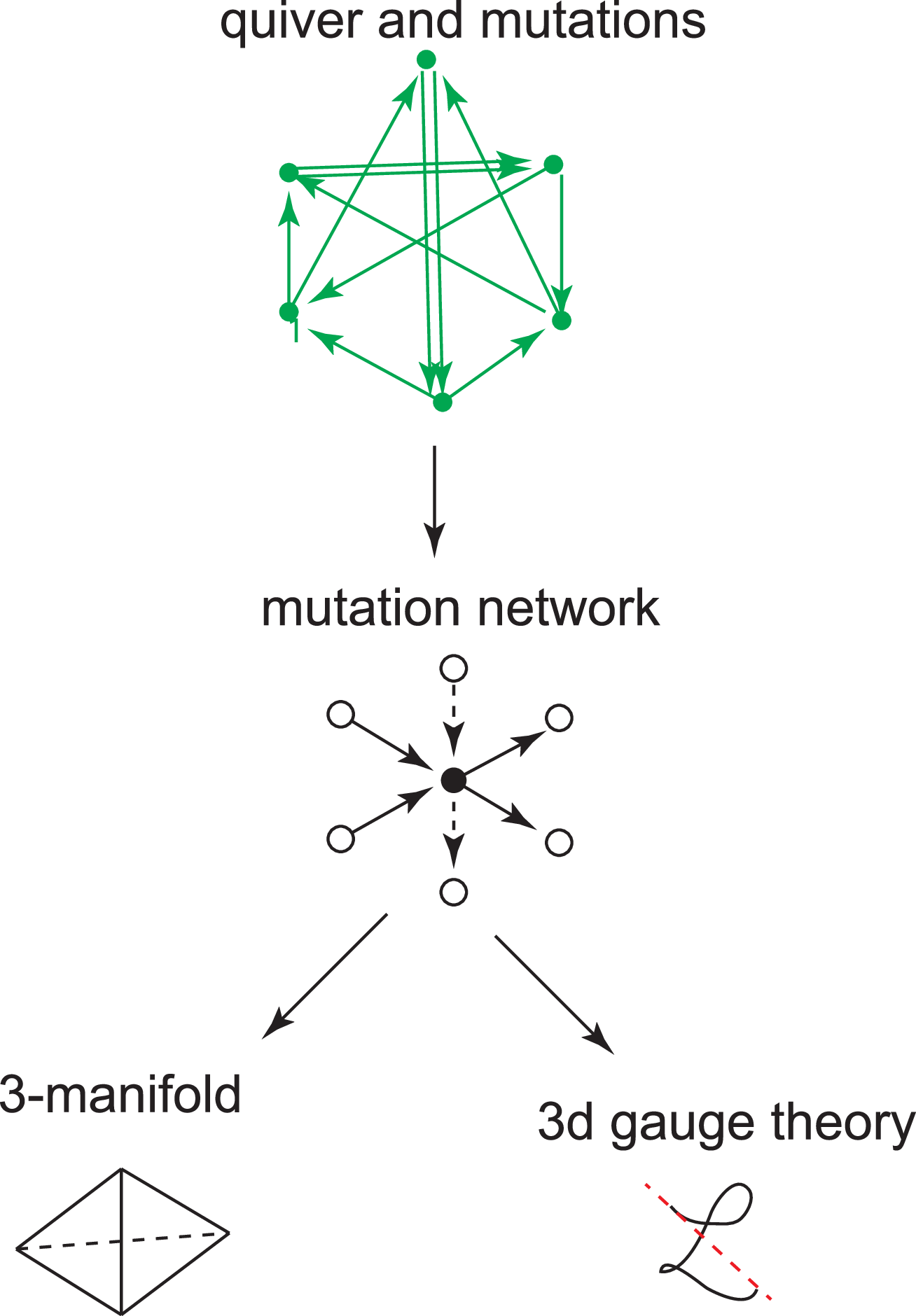}
\caption{Schematic summary of this paper.}
\label{fig.flow}
\end{figure}

The results of this paper are summarized as follows (Figure \ref{fig.flow}):

\begin{enumerate}
\item 
 We introduce a new combinatorial object, a mutation network (section \ref{sec.network}), which encodes the combinatorial data
      of a quiver $Q$ and a mutation sequence $\bm{m}$.

\item We obtain an explicit integral expression for the cluster partition
      function $Z^{\rm cluster}_{(Q,\bm{m})}$ 
      associated with a mutation network (section \ref{sec.derivation}, in particular \eqref{Zmain}).           

\item We outline the construction of a 3d $\scN=2$ theory $\scT[(Q, \bm{m})]$
whose $S^3_b$ partition function coincides with the 
partition function $Z^{\rm cluster}_{(Q,\bm{m})}$ defined previously from the mutation
      network \eqref{ZTQ} (section \ref{sec.identification}).         
 
 \item  In this formulation we find that cluster $x$-variables 
(as opposed to $y$-variables commonly used in the literature in connection with 3-manifolds)
nicely parametrize the global symmetry of the theory.

\end{enumerate}

We also apply our formalism to the quivers and mutations associated with 3-manifolds,
re-deriving and generalizing the previous results
from a unified framework:

\begin{enumerate}
\item When the pair $(Q, \bm{m})$
      satisfies certain conditions, we can construct an associated
      mapping cylinder $(\Sigma, \varphi)$ with a canonical ideal
      triangulation determined from $(Q, \bm{m})$ (section \ref{sec.hyperbolic}).

\item By appropriately identifying boundaries of the
      mapping cylinder $(\Sigma, \varphi)$,
      we obtain a large class of hyperbolic link complements, including arbitrary link
      complements in $S^3$ (section \ref{sec.cap}). 
      This operation has
      a counterpart in the mutation network as well as the partition function.            

\item We consider dimensional reduction of our 3d $\scN=2$ theory to
      2d $\scN=(2,2)$ theory. 
      The twisted superpotential coincides with the Neumann--Zagier
      potential \cite{NeumannZagier} of hyperbolic geometry, 
      and the vacuum equations of
      the 2d $\scN=(2,2)$ theory reproduce the gluing equations 
      of hyperbolic tetrahedra.
\end{enumerate}

Note that our method is different from the 
existing results in the literature (e.g. \cite{Dimofte:2011ju}).
Our approach relies on the Heegaard-like decomposition of 3-manifolds.
This has the advantage of making the connections with the braid group
and the mapping class group 
more direct.
We also point out that cluster $x$-variables, in addition to the cluster $y$-variables
discussed in the literature, play crucial roles in the construction of 3d $\scN=2$ theories.

\bigskip
The rest of this paper is organized as follows.
In section \ref{sec.cluster} we derive an integral expression 
of our partition functions, based on the formalism of quiver mutations
and cluster algebras. The combinatorial data is summarized
in the mutation network. 
We also write down the associated cluster
partition function.
In section \ref{sec.gauge}
we study the 3d $\scN=2$ theory associated with a mutation network.
In section \ref{sec.3mfd} we explain the geometry of hyperbolic
3-manifolds and their ideal triangulations, associated with a mutation
network 
satisfying certain conditions.
The final section (section \ref{sec.conclusion}) summarizes the results
and comments on open problems. 
We include three appendices.
Appendix \ref{sec.dilog} summarizes the 
properties of the quantum dilogarithm function used in the main text.
Appendix \ref{sec.brief} summarizes hyperbolic ideal triangulations and gluing equations.
Appendix \ref{sec.Dehn} explains the effect of the Dehn twist on the
triangulation.

We have tried to make the paper accessible to a wide spectrum of
readers, including mathematicians. In fact, most of the material
in sections \ref{sec.cluster} and \ref{sec.3mfd} (apart from the 
examples in section \ref{sec.example})
require little
prior knowledge of the subject (in physics or in mathematics), and no knowledge of supersymmetric
gauge theories are necessary until section \ref{sec.gauge}.
Readers not interested in 3-manifold cases 
can skip 
section \ref{sec.3mfd}.

%%%%%%%%%%%%%%%%%%%%%%%%%%%%%%%%%%%%%%%%%%%%%
\section{Quivers and Clusters}\label{sec.cluster}
%%%%%%%%%%%%%%%%%%%%%%%%%%%%%%%%%%%%%%%%%%%%%

In this section we define the Hilbert space
$\scH_{Q}$ associated with a quiver $Q$,
 and the action of the mutations $\bm{m}$ 
on $\scH_Q$.
We also define the associated cluster partition function $Z^{\rm cluster}_{(Q, \bm{m})}$,
and derive its integral expression.
This section will be formulated in terms of
\emph{quiver mutations} and \emph{cluster algebras} \cite{FominZelevinsky1}
(see \cite{KellerSurvey} for an introduction).\footnote{For the
appearance of
cluster algebras in 4d $\scN=2$ theories, see for example \cite{Gaiotto:2010be,Cecotti:2010fi}.}

%%%%%%%%%%%%%%%%%%%%%%%%%%%%%%%%%%%%%%%%%%%%%
\subsection{Quiver Mutations and Cluster Algebras}\label{subsec.quiver}
%%%%%%%%%%%%%%%%%%%%%%%%%%%%%%%%%%%%%%%%%%%%%

Let us begin with a \emph{quiver} $Q$, i.e., a finite oriented graph.
We denote the set of the vertices of the quiver by $I$,
and its elements by $i,j, \ldots \in I$.

\begin{figure}[htbp]
  \centering
  \includegraphics[scale=1]{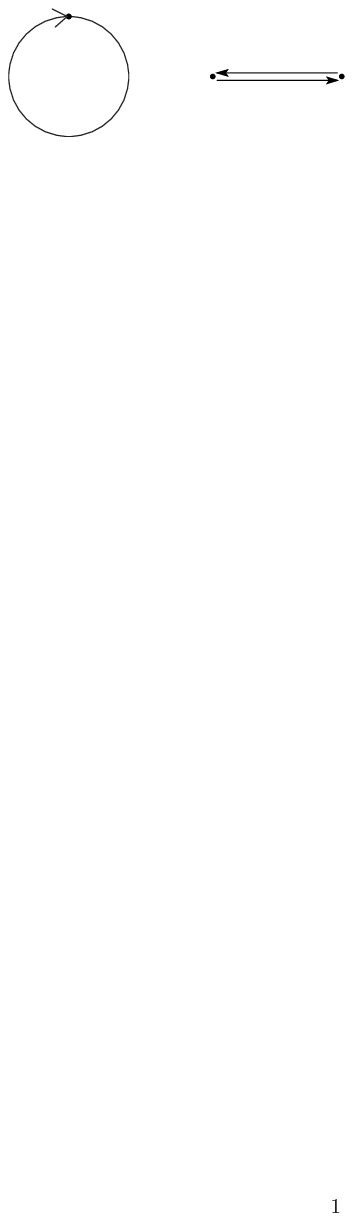}
  \caption{A loop (left) and an oriented $2$-cycle (right) of a quiver.}
\label{fig_loops}
\end{figure}

In this paper, we always assume that a quiver has
no loops and oriented $2$-cycles (see Figure \ref{fig_loops}).

For vertices $i, j\in I$, we define \footnote{$Q_{i,j}$ here is denoted
by $\overline{Q}(i,j)$ in \cite{Nagao:2011aa}.}
\begin{align}
Q_{i,j}:=\#\{\text{arrows from $i$ to $j$}\}-
\#\{\text{arrows from $j$ to $i$}\} \ ,
\end{align}
i.e., $|Q_{i,j}|$ represents the number of arrows from the vertex $i$
to $j$, and the sign represents the chirality (orientation)
of the arrow.
Note that the quiver $Q$ is uniquely determined by the matrix
$Q_{i,j}$ under the assumptions above.

Given a vertex $k$, we define a new quiver $\mu_kQ$ (\emph{mutation} of $Q$
at vertex $k$) by 
\begin{align}
(\mu_kQ)_{i,j}:=
\begin{cases}
-Q_{i,j} &  \text{($i=k$ or $j=k$)} \ , \\
Q_{i,j} + [Q{}_{i,k}]_+[Q_{k,j}]_+ - [Q_{j,k}]_+[Q_{k,i}]_+ &
 \text{$(i,j\neq k$)} \ ,
\end{cases}
\label{Qmutate}
\end{align}
where we used the notation
\begin{align}
[x]_+:=\textrm{max}(x,0) \ .
\end{align}
In more physical language, this can be regarded as a 
somewhat abstract version of the Seiberg duality \cite{Seiberg:1994pq}.
However, the difference here is that our mutations in general
are outside the conformal window.

Let us now construct a non-commutative algebra $\scA_Q$ associated with the quiver
$Q$. This is generated by a set of variables $\sfx_i$,\footnote{In section \ref{sec.identification} we will comment on the 
interpretation of 
these variables as loop operators in 3d $\scN=2$ theories.} associated
with each edge $i$, satisfying the relation\footnote{
More formally, we can think of this space as generated by
$\sfx^{\alpha}$ with $\alpha\in \bZ^{|I|}$, satisfying the relations
\begin{align}
q^{\langle \alpha, \beta \rangle}\sfx^{\alpha} \sfx^{\beta}
=\sfx^{\alpha+\beta}, \quad
\langle \alpha, \beta \rangle =-\langle \beta, \alpha \rangle=^t\!\!\alpha\,
Q \, \beta \ .
\end{align}
In this notation $\sfx_i=\sfx^{e_i}$ for the $i$-th basis of $\bZ^{|I|}$.
}
\begin{align}
\scA_Q:=\{ \sfx_i \, _{(i \in I)} \, |\, \sfx_j \sfx_i=q^{2Q_{i,j}} \sfx_i
\sfx_j \}\ ,
\label{xCCR}
\end{align}
where $q$ is the quantization parameter 
which is related to the parameter $b$ 
by
\beq
q=e^{ \sqrt{-1} \pi b^2 } \ . 
\label{qb}
\eeq
This parameter $b$ will later be identified with the 
deformation parameter of $S^3$ in 
\eqref{ZZ}.
The 
semiclassical limit is given by $q\to 1$, or $b\to 0$.
The variables $\sfx_i$ are the quantum versions
\cite{FockGoncharovEnsembles,FockGoncharovQuantumCluster}
of the 
so-called $y$-variables (coefficients)
in the cluster algebra literature.

We can promote the mutation $\mu_k$ to an operator $\hat{\mu}_k$ on
$\scA_Q$:
\begin{align}
\hat{\mu}_k : \quad \scA_{Q}\to \scA_{\mu_k Q} \ .
\end{align}
The operator $\hat{\mu}_k$ acts on $\sfx_i$ by
\begin{gather}
\sfx'_i:=\hat{\mu}_k (\sfx_i) =
\begin{cases}
\displaystyle
{\sfx_k}{}^{-1} \ ,&(i=k) \ ,\\
\begin{array}{ll}
\displaystyle
\! \! q^{
Q_{ik}[Q_{ki}]_+
}
\sfx_i \sfx_k{}^{[Q_{ki}]_+}
\prod_{m=1}^{|Q_{ki}|}
\big(1+ q^{-\mathrm{sgn}(Q_{ki})(2m-1)}
\sfx_k\big)^{-\mathrm{sgn}({Q_{ki}})} 
\\
\! \!
=
\displaystyle
q^{
Q_{ik}[-Q_{ki}]_+
}
\sfx_i \sfx_k{}^{[-Q_{ki}]_+}
\prod_{m=1}^{|Q_{ki}|}
\big(1+ q^{\mathrm{sgn}(Q_{ki})(2m-1)}
\sfx_k{}^{-1}\big)^{-\mathrm{sgn}({Q_{ki}})} \ .
\end{array}
&
(i\neq k) \ .
\end{cases}
\label{ymutation}
\end{gather}
We can naturally extend the action of $\mu_k$
to the whole of $\scA_Q$.
The resulting variables $\sfx'_i$ satisfy 
commutation relation \eqref{xCCR} for $\mu_k Q$, hence $\hat{\mu}_k$
is indeed a map from $\scA_Q$ to $\scA_{\mu_k Q}$.

The commutation relations \eqref{xCCR},
if written in terms of variables $\mathsf{Y}_i=\log(\sfx_i)$,
take a simple form
\begin{equation}
[\mathsf{Y}_j, \mathsf{Y}_i]=2\pi b^2 \i \,  Q_{i,j}  \ .
\end{equation}
This has a standard representation on a Hilbert space ---
we can choose a polarization, i.e., perform linear transformations 
to find coordinate and 
momentum variables, and coordinates (momenta) act
by multiplication (differentiation).\footnote{In general certain linear
combinations of $\mathsf{Y}_i$ are in the center of the algebra. 
In the case of the quantum \Teichmuller theory discussed in section
\ref{sec.Teichmuller}, the corresponding parameters specify the holonomies
of the flat $SL(2)$ connection at the punctures of the Riemann surface.}
We denote this Hilbert space $\scH_Q$;
the algebra $\scA_Q$ is now the set of operators acting on this state.
We will present more concrete discussion in section \ref{sec.derivation}.

In the following we consider a sequence of quiver mutations
$(\mu_{m_1}, \ldots, \mu_{m_L})$, specified by 
$\bm{m}=(m_1, \ldots, m_L)$ of vertices.
We can think of this as a ``time evolution'' of the quiver,
and for our case at hand will be related to the geometry of Figure
\ref{fig.timeevolution}.
We define the quiver at ``time'' $t$ by
\begin{align}
Q(t):=\hat{\mu}_{m_t} \hat{\mu}_{m_{t-1}}\ldots \hat{\mu}_{m_1} Q\ ,
\quad
Q(0):=Q \ .
\end{align}
We can then define the \emph{cluster partition function} $Z^{\rm cluster}_{(Q, \bm{m})}$ by
\begin{align}
Z^{\rm cluster}_{(Q, \bm{m})}:=\langle \textrm{in} | \hat{\mu}_{m_1} \ldots
\hat{\mu}_{m_L} | \textrm{out} \rangle \ ,
\label{Zeb}
\end{align}
for the initial and final states $|\textrm{in}\rangle \in \scH_{Q(0)}$
and $|\textrm{out} \rangle \in \scH_{Q(L)}$.
The partition function depends on the choice of initial and final
states, whose dependency is suppressed from the notation.

The cluster partition function has been studied in the context of 
the wall-crossing phenomena of 4d $\scN=2$ theories.
The quiver in that context is a BPS quiver 4d $\scN=2$
theories, and the partition function \eqref{Zeb}
is the expectation values of the Kontsevich--Soibelman monodromy
operators \cite{KontsevichSoibelman} (e.g.\ \cite{Cecotti:2010fi});
see also \cite{Cecotti:2011iy,Cordova:2012xk}.

%%%%%%%%%%%%%%%%%%%%%%%%%%%%%%%%%%%%%%%%%%%%%
\subsection{Cluster Partition Functions}\label{sec.derivation}
%%%%%%%%%%%%%%%%%%%%%%%%%%%%%%%%%%%%%%%%%%%%%

Let us next evaluate the partition function \eqref{Zeb}.
In order to convert the operator product in \eqref{Zeb} 
into numbers, we need to 
insert a complete basis set in between the operators,
as is standard in quantum mechanics.
Namely we choose a polarization in $\scH_{Q}$, i.e.\ 
a set of
coordinates $x$ and momenta $p$. Then $\scH_Q$ has
a standard representation on the 
coordinate/momentum basis $|x\rangle, |p\rangle$,
and by inserting a complete basis
\begin{align}
1=\int \! dx \, |x\rangle \langle x | \ ,
\label{completeness}
\end{align}
we can convert the operators into $c$-numbers.

\bigskip

To choose a canonical choice of polarization in $\scH_Q$,
let us prepare a set of variables
$\sfu_i, \sfp_i$ for all the edges $i$, satisfying commutations relations
$[\sfu_i, \sfp_j]=\i \pi b^2 \delta_{i,j}$. Define $\sfx_i$ by
\begin{align}
\sfx_i:=\exp\left( \sfp_i+\sum_j Q_{i,j} \sfu_j \right) \ .
\end{align}
It then follows that the $\sfx_i$s
satisfy the commutation relation in \eqref{xCCR}, justifying the notation.
The variables $\sfu_i, \sfp_i$
have standard representations
on the basis $|u\rangle, |p\rangle$:
\begin{equation}
\begin{split}
&\sfu_i |u\rangle= u_i |u\rangle \ , \quad
\sfp_i |u\rangle=-\i \pi b^2 \frac{\partial }{\partial u_i} |u\rangle  \
,  \\
&\sfu_i |p\rangle=\i \pi b^2 \frac{\partial }{\partial p_i} |p\rangle \ , \quad
\sfp_i |p\rangle=p_i |p\rangle  \ , \quad
\langle u | p \rangle =\exp\left(\frac{1}{\i \pi b^2} u p \right) \ ,
\label{upbasis}
\end{split}
\end{equation}
and moreover we have the completeness
\begin{align}
\int \!du \, |u\rangle \langle u|=1 \ , \quad
\int \!dp \, |p\rangle \langle p|=1 \ ,
\end{align}
with $du:=\prod_i du_i, dp:=\prod_i dp_i$.

Note that this is a highly redundant description of the commutation
relation \eqref{xCCR}; we have doubled the number of variables. 
However the advantage is that 
we can canonically evaluate the expectation values of $\sfx_i$:
\begin{align}
\langle u |  \sfx_i | p \rangle = \exp\left(p_i+\sum_j Q_{i,j} u_j
\right)
\langle u | p\rangle \ . 
\label{uDp}
\end{align}

After these preparations, we could now evaluate the partition function;
the operator $\mu_k$ now acts in a concrete manner in the states
$|u\rangle, |p\rangle$, and we could evaluate its expectation value by 
inserting the complete sets. 
We can go through this exercise 
following \cite{KashaevNakanishi}; see also \cite{Cecotti:2011iy}.\footnote{
In this evaluation we
decompose the action of $\hat{\mu}_k$
into two parts, 
a linear exchange of variables and a conjugation by a quantum
dilogarithm, see
\cite{FockGoncharovQuantumCluster,KashaevNakanishi,Terashima:2011xe}.
In the language of quantum mechanics, this is the translation from the Heisenberg 
picture to 
the Schr\"{o}dinger picture. 
The quantum dilogarithm in \eqref{KNborrow}
comes from the operator representing this conjugation.
}
\footnote{
We need to modify the argument of \cite{KashaevNakanishi} slightly to 
incorporate in and out states. We do not need to 
include their
$\nu^*$, which represents the re-labeling of the edges and is crucial for
the quantum dilogarithm identities of \cite{KashaevNakanishi}
but not for the purposes of this paper.
Note also that $\Phi_b(x)$ in \cite{KashaevNakanishi} is our $e_b(x)^{-1}$.
}
The answer depends on the choice of initial and final states.
Here, we take these to be in the $u$-basis:
\begin{align}
|\,\textrm{in}\rangle=|u(0)\rangle \ ,
\quad
|\,\textrm{out}\rangle=|u(L)\rangle\ . 
\label{ubasis}
\end{align}
We can evaluate the partition function in different initial and final
states 
by converting the expression to the basis of \eqref{ubasis}:
\begin{align}
\langle \textrm{in} | \hat{\mu}_{m_1} \ldots \hat{\mu}_{m_L} | \textrm{out} \rangle =
\int du(0) du(L)\,\,
\langle \textrm{in} | u(0)\rangle
\langle u(0)|  \hat{\mu}_{m_1} \ldots \hat{\mu}_{m_L} | u(L)\rangle 
\langle u(L) | \textrm{out} \rangle \ . 
\label{basischange}
\end{align}
We will come back to the change of initial and final states in section \ref{sec.identification}.
In the basis of \eqref{ubasis}, 
the answer reads
\begin{align}
\begin{split}
Z^{\rm cluster}_{(Q, \bm{m})}
&=\int \prod_{t=0}^{L-1} dp(t) \prod_{t=1}^{L-1} du(t) \prod_{t=0}^{L-1} e_b \left( \frac{1}{2\pi
 b}\left(p_{m_t}(t)-\sum_j Q_{m_t,j}(t)u_j(t)\right) \right)^{-1} \\
& \qquad \qquad 
\times \exp \left( \frac{\i}{\pi b^2} (u(t)p(t)-u(t+1)\tilde{p}(t+1))
 \right) 
\ ,
\end{split}
\label{KNborrow}
\end{align}
where
\begin{align}
\tilde{p}_i(t+1):=
\begin{cases}
-p_{m_t}(t)   &  (i=m_t) \ , \\
p_i(t) +[Q_{m_t,i}(t)]_+ p_{m_t}(t) &  (i\ne m_t)  \ ,
\end{cases} 
\end{align}
and $e_b(z)$ is the quantum dilogarithm function
defined in Appendix \ref{sec.dilog}. 
For notational simplicity we did not explicitly show some of the indices $i,j, \ldots \in I$; for
example $u(t) p(t):=\sum_i u_i(t) p_i(t)$.

This expression \eqref{KNborrow} was obtained in \cite{KashaevNakanishi}. 
Our observation is that one can rewrite this expression
into a form more suitable for the identification of 
3d $\scN=2$ theory.

To explain this, first note that the expression \eqref{KNborrow}
has a large number of integral variables $u_i(t), p_i(t)$.
Most of them can be trivially integrated out.
Indeed, the power of the exponent in \eqref{KNborrow} 
reads
\begin{align*}
u(t)p(t)-u(t+1) \tilde{p} (t+1)&=
u_{m_t}(t) p_{m_t}(t)+\sum_{i\not= m_t} u_i(t) p_i(t) \\
&  +u_{m_t+1}(t+1) p_{m_t}(t)-\sum_{i\not= m_t} u_i(t+1)(p_i(t)+
[ Q_{m_t,i}(t) ]_+ p_{m_t}(t)) \ .
\end{align*}
In particular the variable $p_i(t)$ $(i\ne m_t)$ does not appear inside the argument of 
the dilogarithm, and only appears in the linear term
$(u_i(t)-u_i(t+1)) p_i(t)$.
Integrating out $p_i(t) \ \ (i \not= m_t)$, we have
\begin{align}
u_i(t)=u_i(t+1) \qquad (i \not= m_t) \ ,
\label{uconstraint}
\end{align}
and hence most of the $u$-variables are identified,
leading to
\begin{align}
\begin{split}
Z^{\rm cluster}_{(Q, \bm{m})}
&=\int \prod_{t=0}^{L-1} dp_{m_t}(t) \prod_{t=1}^{L-1} du(t) \prod_{t=0}^{L-1} e_b \left( \frac{1}{2\pi b} \left(p_{m_t}(t)-\sum_j Q_{m_t,j}(t)u_j(t)\right) \right)^{-1}\\
& \quad \quad \quad \times \exp \left( \frac{\i}{\pi b^2} p_{m_t}(t)
 (u_{m_t}(t)+u_{m_t}(t+1)-\sum_{i\not= m_t} [ Q_{m_t, i}(t) ]_+u_i (t))
 \right) 
\ .
\end{split}
\end{align}
We can integrate over $p_{m_t}(t)$ by \eqref{ebFourier} in Appendix A,
leading to (up to a constant overall phases irrelevant for the identification of 3d $\scN=2$ theories)
\begin{align}
\begin{split}
Z^{\rm cluster}_{(Q,\bm{m})}=
\int \prod_{t=1}^{L-1} du(t)
\,\, \prod_{t=0}^{L-1} e_b\left(\frac{1}{2 \pi b} Z'(t)
 -\frac{i\scQ}{2}\right)^{-1} 
\exp\left(-\frac{\i\pi}{(2 \pi b)^2}
Z'(t) Z''(t)
\right) \ ,
\end{split}
\label{Zresult}
\end{align}
where we defined $Z'(t), Z''(t)$ by
\begin{align}
\begin{split}
Z'(t)&:=2\left[-u_{m_t}(t)-u_{m_t}(t+1)+\sum_{i} [Q_{m_t,
i}(t)]_+ u_i(t) \right]\ , \\
Z''(t)&:=2\left[u_{m_t}(t)+u_{m_t}(t+1)-\sum_{i} [-Q_{m_t,
i}(t)]_+ u_i(t) \right] \ ,
\label{crossratio}
\end{split} 
\end{align}
and
\begin{align}
\scQ:=b+b^{-1} \ .
\end{align}
For later purposes we also define
\begin{align}
Z(t)&:=\i\pi b\scQ -2\left[
\sum_{i} [Q_{m_t, i}(t)]_+ u_i(t) 
-\sum_{i} [-Q_{m_t,
i}(t)]_+ u_i(t)
\right] \nonumber \\
&=
\i\pi b\scQ-2\left(
\sum_{i} Q_{m_t, i}(t) u_i(t) 
\right)
  ,
\label{Zt}
\end{align}
satisfying 
\beq
Z(t)+Z'(t)+Z''(t)=\i\pi b\scQ \  .
\eeq
The factor $\i\pi b \scQ$ will turn out to be useful for later considerations.

In \eqref{Zresult},
we need to remember
that the variables $u_i(t)$ are constrained by \eqref{uconstraint}.
To take these constraints into to account,
it is useful to introduce a notion of a \emph{mutation network}.

%%%%%%%%%%%%%%%%%%%%%%%%%%%%%%%%%%%%%%%%%%%%%
\subsection{Mutation Networks}\label{sec.network}
%%%%%%%%%%%%%%%%%%%%%%%%%%%%%%%%%%%%%%%%%%%%%

\begin{figure}[htbp]
\centering{\scalebox{0.9}{\includegraphics{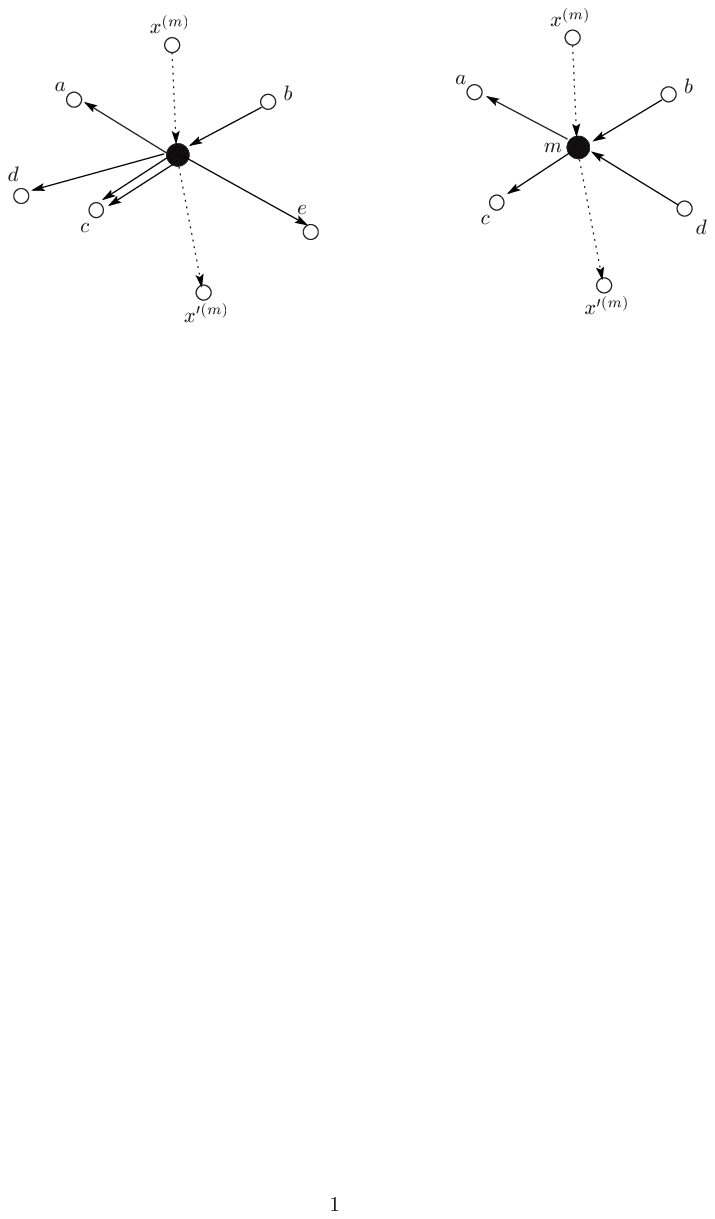}}}
\caption{(Left) A mutation network around a mutation $m$.
The mutation is represented by a black vertex $m$,
and the integral variables are represented by 
white vertices.
Two of the white vertices, $x^{(m)}$ and  $x'^{(m)}$,
represent the variables
associated with the mutated vertices,
and are connected by dotted lines.
Other white vertices are connected by undotted lines, and 
come with the charges $Q_{m,w}$.
(Right) When the quiver arises from an ideal triangulation of a 2d
 surface
(section \ref{sec.Teichmuller}), a mutation network always looks as in
 this figure around a black vertex.
}
\label{aroundwhite}
\end{figure}

Let us again begin with a pair $(Q, \bm{m})$,
a quiver $Q$, and sequence of mutations
$\bm{m}=\{ m_1,m_2,...,m_L\}$.
We then associate a graph, a mutation network
(see Figure 
\ref{aroundwhite}). 

The graph is always bipartite, i.e., 
vertices are colored either black or white, 
and edges connect
vertices of different colors.
We denote the set of black (white) vertices of the network by $B$ ($W$).

A black vertex represents a mutation, one of the $\bm{m}$'s.
A white vertex represents the integral variables $u_i(t)$
of the previous subsection.

The black vertex $m\in B$, representing a mutation $m$,
 is connected to two white vertices 
(denoted by $x^{(m)}, x'^{(m)}$ in Figure \ref{aroundwhite})
by dotted lines --- these represent the variables $u_{m_t}(t)$
and $u_{m_t}(t+1)$ in the previous subsection.
A black vertex is also connected with other white vertices by 
undotted lines --- these white vertices represent 
the $u_i(t)$s with $i\ne m_t$.
The number of undotted arrows from $m\in B$ to $w\in W$
is determined by $Q_{m,w}:=Q(t)_{m_t,w}$,
which is one of the components of the quiver adjacency matrix
\emph{before} the mutation.
In general there are multiple arrows from $m$ to $w$ (or from $w$ to $m$,
 depending on the sign of $Q_{m,w}$).

This rule defines the mutation network.
Around a black vertex, the network represents the mutation and the 
quiver vertices affected by it. Around a white vertex,
the network describes the creation of a new integral at some
time, and its annihilation at a later time.
The network is naturally concatenated when we combine two mutation sequences.
Concrete examples of mutation networks 
will appear later in section \ref{sec.example}.

\bigskip

Let us come back to our partition function \eqref{Zresult}.
We learn from \eqref{uconstraint}
that the independent variables are associated with the white vertices of the
mutation network; we denote this variable by $x_w\, (w\in
W)$.
Some of the edges are in the initial (final) quiver $Q(0)$
($Q(L)$), and others not; we denote the difference by $w\in W_{\rm ext}$
if they are, and $w\in W_{\rm int}$ if they are not.
By definition we have 
$$
W_{\rm ext}\cup W_{\rm int}=W, 
\quad
W_{\rm ext}\cap
W_{\rm int}=\emptyset \ .
$$

We can now rewrite the result \eqref{Zresult} as
\begin{align}
Z^{\rm cluster}_{(Q, \bm{m})}=\int \prod_{w\in W_{\rm int}}dx_w \,\,  \prod_{m\in B}  
e_b\left(\frac{1}{2 \pi b} Z'(m)-\frac{\i\scQ}{2}\right)^{-1}
\, 
\exp\left(-\frac{\i\pi}{(2 \pi b)^2} Z'(m) Z''(m) \right)  
\ ,
\label{Zmain}
\end{align}
where we defined
\begin{align}
\begin{split}
Z'(m)&:=2\left[-x^{(m)}-x'^{(m)}+\sum_{w\in W} [Q_{m,
w}]_+  x_w \right]\ , \\
Z''(m)&:=2\left[x^{(m)}+x'^{(m)}-
\sum_{w\in W} [-Q_{m,w}]_+ x_w \right] \ ,
\end{split} 
\end{align}
and
\begin{equation}
\begin{split}
Z(m):&=\i\pi b\scQ-2\left[
\sum_{w\in W} [Q_{m, w}]_+ x_w
-\sum_{w\in W} [-Q_{m,
w}]_+ x_w
\right]
\\
&=
\i\pi b\scQ-
2\left(
\sum_{w\in W} Q_{m, w}\, x_w
\right)\ . 
\end{split}
\end{equation}
Note that the integrand of \eqref{Zmain}
factorizes into contributions from 
each mutation ($m$). 
The expression \eqref{Zmain} will be the crucial ingredient for the 
identification of 3d gauge theories in section \ref{sec.identification}.

\bigskip

When we discuss 3-manifolds (and associated 3d $\scN=2$ theories), 
we concentrate on the case when the quiver 
is determined from an ideal triangulation of a Riemann surface
(section \ref{sec.Teichmuller}). 
In this case the mutation network always looks as in the right of Figure
\ref{aroundwhite}
around a white vertex, namely two lines corresponding to mutated
vertices, and four lines with two charge $+1$ (denoted by 
$a^{(m)}, c^{(m)}$) and two $-1$ (denoted by $b^{(m)}, 
d^{(m)}$).\footnote{$a^{(m)}$ and $c^{(m)}$, or $b^{(m)}$ and $d^{(m)}$, could
be identified with each other, as in the case of the quiver coming from the triangulation of 
the once-punctured torus.} 
In this notation we have
\begin{align}
\begin{split}
Z(m)&=\i\pi b\scQ+2\left(- a^{(m)}-c^{(m)}+b^{(m)}+d^{(m)}\right) \ , \\
%\! \!\quad
Z'(m)&=2\left(-x^{(m)}-x'^{(m)}+a^{(m)}+c^{(m)} \right) \ , \\
Z''(m)&=2\left( x^{(m)}+x'^{(m)}-b^{(m)}-d^{(m)}\right) \ ,
\label{ZZ'}
\end{split}
\end{align}
where for notational simplicity we used $a^{(m)}$ also for 
the associated variable, which should be written $x_{a^{(m)}}$ 
in our previous notation.
Interestingly, this (in particular the variable $Z(m)$) is precisely
the coordinate transformation from
the cluster $x$-variables to the cluster $y$-variables,
or in the \Teichmuller language from the Penner coordinates (geodesic
length)
to the 
Fock coordinates (shear coordinates, i.e. cross ratios); see 
Table \ref{table.dict}.\footnote{A numerical factor of $2$ in \eqref{ZZ'}
is consistent with \cite[Remark 5.1]{KashaevNakanishi}.
} 

\begin{figure}[htbp]
\centering{\includegraphics[scale=1]{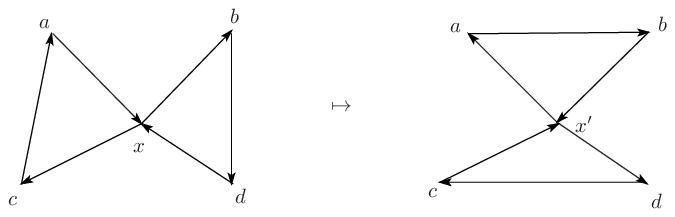}}
\caption{Mutation of a quiver at vertex $x$, representing a flip of an
 2d ideal triangulation.}
\label{mutatex}
\end{figure}

\begin{table}[htbp]
\caption{Dictionary between mutation networks,
\Teichmuller theory and cluster algebras. The \Teichmuller
 interpretation does not exist for general quivers which do not arise
 from an ideal triangulation of a 2d surface.}
\label{table.dict}
\bigskip
{\renewcommand\arraystretch{1.4}
\begin{center}
\begin{tabular}{c|c|c}
mutation network & cluster algebra & \Teichmuller space \\
\hline
\hline
variable $x_w \,\, (w\in W)$ & cluster $x$-variable & Penner coordinate  \\
\hline
variable $Z(m)\,\,  (m\in B)$ & cluster $y$-variable & Fock coordinate \\
\end{tabular}
\end{center}
}
\end{table}

\bigskip
Coming back to the general expression \eqref{Zmain},
we find that the integrand depend on $2|B|$ variables $Z'(m), Z''(m)$.
As we mutate the quiver, the number of such variables 
grows at roughly twice the speed of the number of variables $x_w$'s 
($|W_{\rm int}|$), since we obtain two new variables 
$Z'(m), Z''(m)$ for each mutation.\footnote{In 
general we have the relation
\beq
|W_{\rm int}|+|W_{\rm ext}|=|W|=|I |+ |B| \ .
\eeq
}
This means that there should be constraints along $Z'(m), Z''(m)$'s.
In fact, we find the following $|W_{\rm int}|$ constraints.

Suppose that we fix a white 
internal vertex $w\in W_{\rm int}$. 
This vertex is connected with 
many mutations $m\in B$, and two of mutations are connected with
$w$ by dotted lines --- a vertex is created by a mutation, and deleted by a mutation after some steps.
We call these mutations \emph{type 1}.
The remaining mutations are called \emph{type 2} (of \emph{type 3})
if $Q_{m,w}<0$ (if $Q_{m,w}>$0).
We then find 
\beq
\sum_{m:\ {\rm type 1}} Z(m)+\sum_{m:\ {\rm type 2} } [-Q_{m,w}]_+ Z'(m)+\sum_{m:\ {\rm type 3}} [Q_{m,w}]_+ Z''(m)=2\pi b\scQ  \i\ .
\label{sumZ}
\eeq
This relation will 
correspond to the superpotential constraints for R-charges in 
3d $\scN=2$ theories in section \ref{sec.identification}.
 
To show \eqref{sumZ}, it is useful to go back to the notation of \eqref{Zresult}, 
under which \eqref{sumZ} reads
\beq
Z(t_1)+Z(t_2)+\sum_{t_1<t<t_2 } [-Q_{m_t,i}]_+ Z'(t)+\sum_{t_1<t<t_2} [Q_{m_t,i}]_+ Z''(t)=2\pi b\scQ  \i\ ,
\label{sumZ2}
\eeq
where we used the notation that the variable $u_i(t)$, corresponding to $w$, 
 is generated at $t=t_1$ and annihilated at $t=t_2$, i.e.,
 $i=m_{t_1}=m_{t_2}$.
Using the definitions \eqref{crossratio} and \eqref{Zt},
the left-hand side of \eqref{sumZ2} becomes
\begin{align}
2\left[
-\left(
\sum_j Q(t_1)_{m_{t_1},j} u_j(t_1) 
+\sum_j  Q(t_2)_{m_{t_2},j} u_j(t_2) 
\right)
+ \sum_{t_1<t<t_2} Q(t)_{m_t,i} \left(u_{m_t}(t)+u_{m_t}(t+1)\right) \right.
\nonumber\\
\left.
+\sum_{t_1<t<t_2} \sum_j  \left(
[Q(t)_{i,m_t}]_+ [Q(t)_{m_t,j}]_+
- 
[Q(t)_{j,m_t}]_+ [Q(t)_{m_t,i}]_+
\right)u_j(t)
+\pi b\scQ\i
\right] \ .
\label{tmp1}
\end{align}
From the definition of mutation \eqref{Qmutate}
and the constraints on the variables $u_i(t)$ \eqref{uconstraint}, 
the third term inside the bracket of \eqref{tmp1} is equivalent to
\begin{align*}
\sum_{t_1<t<t_2} \sum_{j\ne m_t}  \left(
Q(t+1)_{i,j}-Q(t)_{i,j}
\right)u_j(t)
=
\sum_{t_1<t<t_2} \sum_{j\ne m_t} \left(
Q(t+1)_{i,j}u_j(t+1)-Q(t)_{i,j}u_j(t) 
\right)   \ .
\end{align*}
The sum over $j$ becomes over all $j\in I$
when this is combined with the second term in \eqref{tmp1}.
After many cancellations, this leads to
\begin{align*}
\sum_j \left(
Q(t_2)_{i,j} u_j(t_2)- Q(t_1+1)_{i,j}u_j(t_1+1)
\right) 
=\sum_j \left(
Q(t_2)_{i,j}u_j(t_2)+Q(t_1)_{i,j}u_j(t_1) 
\right) \ .
\end{align*}
This cancels the first term in \eqref{tmp1},
with the only remaining term in \eqref{tmp1} being the fourth term, 
giving $2\pi b \scQ \i$. This proves \eqref{sumZ}.

%%%%%%%%%%%%%%%%%%%%%%%%%%%%%%%%%%%%%%%%%%%%%
\section{%\texorpdfstring
{3d $\scN=2$}
%{3d N=2} 
Theories}\label{sec.gauge}
%%%%%%%%%%%%%%%%%%%%%%%%%%%%%%%%%%%%%%%%%%%%%

In this section we outline the construction of 3d $\scN=2$ theories
$\scT[(Q,\bm{m})]$
satisfying \eqref{ZTQ}, based on the results of 
section \ref{sec.cluster}.

\subsection{
%\texorpdfstring
{$S^3_b$}
%{S3} 
Partition Functions}

Let us first summarize the basic ingredients of 3d $\scN=2$ theories,
and their $S^3_b$ partition functions \cite{Kapustin:2009kz,Jafferis:2010un,Hama:2010av,Hama:2011ea}.

Our theories have a number of $U(1)$ symmetries, which are labeled by the set $H$.
Some of the $U(1)$ symmetries are flavor symmetries, and others gauge
symmetries.
We denote this by $i\in F$ and $i\in G$, respectively.
By definition we have $F\cup G=H$, and $F\cap G=\emptyset$.
For each $i$ there is a corresponding parameter $\sigma_i$, the scalar of the
associated $\scN=2$ vector multiplet. When $i\in F$, the correspondingly vector multiplet is 
non-dynamical
and the parameter $\sigma_i$ is called 
a real mass parameter.

We also include Chern--Simons terms, including off-diagonal ones.
This could be described by a symmetric matrix $k_{i,j}$ for $i,j\in H$:
\begin{equation}
\sum_{i,j \in H}\frac{k_{ij}}{4\pi}\int A_i\wedge d A_j \ ,
\end{equation}
for dynamical/background gauge field $A_{i,j}$.
The Chern--Simons term has to obey a quantization condition
for the invariance under the large coordinate transformation 
and for the absence of parity anomaly,
and in particular $k_{i,j}$ should be a half-integer.
Note also that we consider $k_{i,j}$
for either $i$ or $j$ in $F$; these are Chern--Simons terms for background 
Chern--Simons terms, which play crucial roles when we gauge the associated
global symmetry.

The $S^3_b$ partition function of a 3d $\scN=2$ theory 
depends on 
the Chern--Simons term $k_{i,j} \, (i,j\in H)$
and the real mass
%/FI 
parameters $\sigma_i \, (i\in F)$.
Here $S^3_b$ is a 1-parameter deformation of the $S^3$-preserving
$U(1)\times U(1)$ isometry. More explicitly it is defined by \cite{Hama:2011ea}
\begin{align}
b^2 |z_1|^2+b^{-2}|z_2|=r^2 \ ,
\end{align}
with $z_1, z_2\in \bC$. The partition function turns out to be
independent of $r$.

Now the $S^3_b$ partition function, after localization computation,
takes the following form:
\begin{align}
Z\left[k_{i,j \,\, (i,j\in H)} , \sigma_{i\,\, (i\in F)}\right]
=\int \left( \prod_{i\in G}  d\sigma_i 
\right) Z_{\rm classical}(k, \sigma)\, Z_{\rm 1-loop}(\sigma) \ .
\label{ZS3}
\end{align}
The rules are summarized as follows (we specialize to Abelian gauge
theories in this paper):
\begin{itemize}
\item

The integral is over all the Abelian global symmetries $\sigma_i$, $(i\in
     G)$.
\item

The classical contribution $Z_{\rm classical}(k, \sigma)$ is determined
     by the Chern--Simons term,\footnote{There are also classical
     contributions from FI parameters.} and
     is given by\footnote{$k_{i,j}$ here is the bare Chern--Simons term,
     not the effective Chern--Simons term obtained after integrating out
     massive matters.}
\begin{align}
Z_{\rm classical}(k,\sigma)=\exp\left(-\i\pi \sum_{i,j\in H} 
k_{i,j} \sigma_i \sigma_j\right) \ .
\end{align}

\item 

The 1-loop determinant $Z_{\rm 1-loop}$ has contributions only from
      $\scN=2$ chiral multiplets.
This is given by
\begin{align}
Z_{\rm 1-loop}(\sigma)=\prod_{w: \, {\rm hyper}}s_b\left(\sum_{i \in H}
      Q_{i, m} \sigma_i+\frac{\i\scQ}{2}(1-q_m)\right)\ ,
\label{Z1loop}
\end{align}
where we assumed the $\scN=2$ chiral multiplet $m$ has charges $Q_{i,m}$ under the
      $U(1)_i$ symmetry, and has R-charge $q_m$.
The correct value of the IR superconformal R-charge
is dependent on the mixing of the UV $U(1)$ R-symmetry with flavor symmetries \footnote{The correct mixing is determined by $F$-maximization \cite{Jafferis:2010un}.}, and we can write \eqref{Z1loop} as the holomorphic combination\footnote{The reason
why the Coulomb branch parameter (real mass parameter)
and the R-charge appear in a holomorphic combination has been 
clarified in \cite{Festuccia:2011ws} in the background supergravity formalism.
}
\begin{align}
Z_{\rm 1-loop}(\sigma)=\prod_{m: \, {\rm hyper}}s_b\left(\frac{\i\scQ}{2}-\sum_{i \in H}Q_{i,m} \tilde{\sigma_i} \right)\ ,
\label{Z1loop2}
\end{align}
with 
\beq
\textrm{Re}(\tilde{\sigma}_i)=\sigma_i, \quad 
\sum_{i\in H} Q_{i,m}\, \textrm{Im}(\tilde{\sigma}_i)=\frac{\scQ}{2} \, q_m \ .
\label{sigmareim}
\eeq

\end{itemize}

Note that in the partition function \eqref{ZS3} the only distinction between $\sigma_i$ for $i\in F$
     and
those for $i\in G$ is that we do not integrate over the
     former. This means that effect of gauging a global symmetry is simply to
     integrate over the corresponding $\sigma_i$ in the $S^3_b$ partition function.

%%%%%%%%%%%%%%%%%%%%%%%%%%%%%%%%%%%%%%%%%%%%%
\subsection{Properties of 3d 
%\texorpdfstring
{$\scN=2$}
%{N=2} 
Theories} \label{sec.identification}
%%%%%%%%%%%%%%%%%%%%%%%%%%%%%%%%%%%%%%%%%%%%%

Let us finally comment on the properties of our theories
$\scT[(Q,m)]$. We will content ourselves with  
comments on the basic properties and 
defer the detailed analysis of our theories
to a future publication.

The theory $\scT[(Q,m)]$ in this paper is defined 
in such a way that the relation \eqref{ZTQ} holds:
the $S^3_b$ partition function
\eqref{ZS3} should be compared 
with our partition function \eqref{Zmain}, which we rewrite
here (using \eqref{ebsb} and \eqref{sbinverse}) 
to be
\begin{align}
Z^{\rm cluster}_{(Q, \bm{m})}=\int \prod_{w\in W}dw\,\,  \prod_{m\in B} 
s_b\left(\frac{\i \scQ}{2}-\tsigma_m\right)
\, 
\exp\left(-\frac{\i\pi}{2} \left(\frac{\i \scQ}{2}-\tsigma_m
 \right)^2- \i \pi  \tilde{\sigma}_m \tilde{\sigma}'_m
\right)  \ , 
\label{Zmain2}
\end{align}
where we defined
\beq
\tsigma_m: =\frac{1}{2\pi b}Z'(m), \quad
\tilde{\sigma}'_m: =\frac{1}{2\pi b}Z''(m) \ ,
\label{sigmaq}
\eeq
and we have again neglected overall phase factors.

The properties of the theory $\scT[(Q,\bm{m})]$ are summarized as follows.

\subsubsection*{1. $\scN=2$ Abelian Vector Multiplets ($U(1)$ Symmetries)\\ }

We associate a $U(1)$ symmetry for each white vertex of the
      mutation network. This is either a global or gauge symmetry 
depending on whether or not the white vertex is associated with the
      quiver edge in the initial/final states.
\beq
F=W_{\rm ext} \ , \quad G=W_{\rm int} \ , \quad 
H=W \ .
\eeq
We also identify the corresponding variables
$x_w$ and $\tilde{\sigma}_i$ by the relation
\beq
\tilde{\sigma}_i=\frac{1}{2\pi b} 2 x_w \ .
\label{sigmaixw}
\eeq 
For the case of 3-manifolds, an equivalent way to say this is that the
      $U(1)$
symmetry is associated with an edge of the tetrahedron, 
and is a global (gauge) symmetry if the edge is on the boundary (in the
      interior) of the 3-manifold.

Note that not all the $U(1)$ symmetries are really 
independent --- in fact, many of them are related by 
electric--magnetic duality, and this happens when the corresponding variables $\sfx_i$ do not commute.
Note also that the $U(1)$ symmetries in general have Chern--Simons terms,
and are determined by the quadratic expression in \eqref{Zmain2}.

\subsubsection*{2. $\scN=2$ Chiral multiplets \\} 

We associate an $\scN=2$
chiral multiplet $\Phi_m$
for each black vertex $m\in B$ of the
      mutation network. The charges $Q_{m,w}$ associated with the edges of the mutation network 
determine the charges of these fields under the $U(1)$ symmetries;
the parameter $\tilde{\sigma}_m$, which appears inside the argument of
$s_b$, is a linear combination of $\tilde{\sigma}_i$'s due to the relations
 \eqref{crossratio}, \eqref{sigmaq} and \eqref{sigmaixw}.
In particular the R-charge of $\Phi_m$ is given by
\beq
\frac{\scQ}{2} q_m= \textrm{Im}(\tilde{\sigma}_m)=
\textrm{Im}\left(\frac{Z'(m)}{2\pi b}\right) \ .
\eeq
For the case of 3-manifolds, this chiral multiplet is associated with an
      ideal tetrahedron.

\subsubsection*{3. Superpotential \\} Finally, there is a superpotential
term among the $\scN=2$
      chiral multiplets.
Given a white vertex in the mutation network, we associate a
      superpotential term.
Depending on the three types (types 1, 2, 3 discussed in section
\ref{sec.network}),
the operator $\scO_{m,w}$ is either a fundamental field or 
involves monopole operators.
This means that our theories are generically non-Lagrangian.
The fact that the superpotential has R-charges $2$ 
could be guaranteed by the 
relation \eqref{sumZ} we have proven earlier.

\subsubsection*{4. The $Sp(2N, \bZ)$ Action \\}

Let us come back to the choice of initial and 
final states of the partition function \eqref{Zeb}; our discussion
so far assumes the choice \eqref{ubasis}.

As we discussed in \eqref{basischange}, a change of the boundary
condition induces a change of the partition function.
For example, when we change $\langle\textrm{in}|$
from $\langle u(0)|$ to $\langle p(0)|$,
we have
\begin{align}
\langle p(0)| \hat{\mu}_{m_1}\ldots \hat{\mu_L} |u(L)\rangle
=\int du(0) \, \, e^{\frac{1}{\pi b^2\i} u(0) p(0)} 
\, \langle u(0)| \hat{\mu}_{m_1}\ldots \hat{\mu_L} |u(L)\rangle
\ ,
\label{ZFourier}
\end{align}
which is just a Fourier transformation.
More generally, we could choose a different state by
an $Sp(2N, \bZ)$-transformation, where $N$ here is given by $2|I|$
(note that we could choose to mix variables $u_i(0), p_i(0)$ and $u_i(L),
p_i(L)$).
This $Sp(2N, \bZ)$
action is lifted to the action of the wavefunction,
which in turn is identified with the $Sp(2N, \bZ)$ action on 
3d $\scN=2$ Abelian theories \cite{Witten:2003ya}.
This involves adding Chern--Simons terms for background gauge fields
and gauging global symmetries with off-diagonal Chern--Simons 
terms.

\subsubsection*{5. Gauging \\}
Gluing two 3d $\scN=2$ theories is represented by 
a concatenation of the mutation network.
When we have several networks $\scN_i$ and glue them together at
      a white
vertex $w$, 
we gauge the diagonal subgroup of 
corresponding global symmetries $U(1)^M$, and
we have
\beq
Z_{\cup_w \scN_i}=\int dx_w\,\, \prod_i  Z_{\scN_i}(x_w)\ ,
\eeq
where we showed the dependencies of $Z_{\scN_i}$ only with respect $x_w$.
Such a gauging is necessary, for example, when we cap off the braids in
section \ref{sec.cap}.

\subsubsection*{6. Dimensional Reduction and SUSY Moduli Space \\}
The semiclassical limit $b\to 0$ discussed in section
 \ref{sec.structure}
is the limit where the ellipsoid
$S^3_b$ degenerates into $\bR^2\times S^1_b$ with a circle of small
radius $b$.
Since we take $b$ to be small we are effectively reduced
to 2d $\scN=(2,2)$ theory, but with all the KK modes taken into account.
The parameters $x_w$ will play the role of the vector multiplet scalars
(which is complexified after dimensional reduction), 
and $\scW$ is the effective twisted superpotential
obtained by integrating out matters.
The gluing equation then is the 
vacuum equation for the 2d theory (cf.\ \cite{Nekrasov:2009uh}). 
The SUSY 
moduli space of our 3d $\scN=2$ theory 
should be constructed from the symplectic quotient construction 
(cf.\ \cite{Dimofte:2011gm}), imposing \eqref{sumZ} as constraints.

\subsubsection*{7. Loop Operators \\}
We propose that the variables $\sfx_i$ represent the flavor Wilson/vortex operators
for the $i$-th flavor symmetry. In fact, flavor Wilson (and vortex) loop
operators
wrapping the Hopf fiber at the north pole\footnote{We can also consider
loop operators located at the south poles, and the corresponding
$y$-variables. This realizes the $b\leftrightarrow 1/b$ symmetry of our theory.} of the base $S^2$
represent the multiplication and shift to the partition function \cite{Kapustin:2012iw,Drukker:2012sr}, 
leading to the commutation relation \eqref{xCCR}.
This means that insertion of the $\sfx_i$ in the cluster partition function \eqref{Zeb} should
be identified with the (unnormalized) VEV of the corresponding loop operator.
The situation is similar to the case of
3d $\scN=2$ theories 
coming from the dimensional reduction of 4d $\scN=1$ theories in
\cite{Heckman:2012jh}; that paper also proposes identifying
(classical) $y$-variables with 
the VEVs of loop operators.

\subsubsection*{8. Coupling to 4d $\scN=2$ Theory \\}

Our quiver in many cases can be identified with a BPS quiver for a
4d $\scN=2$ theory. In these cases it is natural to propose that 
our 3d theory arises on the boundary of the 4d $\scN=2$ theory;
the latter couples to the former by gauging global symmetries of the former.
The BPS wall crossing causes the mutation of the BPS quiver,
which is translated into the change of duality frames
of our 3d $\scN=2$ theories (see \cite{Cecotti:2011iy} for a physical explanation of this 
correspondence, in the case associated with 3-manifolds).
This includes the case where 4d $\scN=2$ theory is complete in the 
classification of  \cite{Cecotti:2011rv}, and (in addition to several exceptional cases)
the quivers coming from the triangulation, discussed in section \ref{sec.Teichmuller}.
Our analysis suggests that this correspondence is more general,
and holds for non-complete 4d $\scN=2$ theories;
the classification program of the IR fixed points of 3d $\scN=2$ theories 
is closely related with the classification of 
4d $\scN=2$ theories! 

%%%%%%%%%%%%%%%%%%%%%%%%%%%%%%%%%%%%%%%%%%%%%%%%%%%
\section{3-manifolds}\label{sec.3mfd}
%%%%%%%%%%%%%%%%%%%%%%%%%%%%%%%%%%%%%%%%%%%%%%%%%%%

In this section we apply the formalism of the previous section to 
the special case of 3d $\scN=2$ theories associated with hyperbolic
3-manifolds.

In this case, the quiver is determined from an ideal triangulation of
a 2d surface,
and the mutation sequence represents the action of the  
mapping class group.
The Hilbert space then is that of the quantum \Teichmuller space.

The goal of this section is threefold.
We first discuss the canonical ideal triangulation of our 3-manifold
(section \ref{sec.hyperbolic}), which originates from an ideal triangulation of a 
2d surface. Second, we discuss how to modify the construction to 
obtain more general geometries, by identifying unglued faces of a
mapping cylinder (section \ref{sec.cap}).
This gives new methods to systematically study the geometry of link complements,
and the results of the previous sections automatically gives
associated 3d $\scN=2$ theories.
Third, we discuss the gluing equations of hyperbolic tetrahedra, 
and show that it arises from the semiclassical limit of the partition
function
discussed in section \ref{sec.cluster} (section \ref{sec.structure}). 

%%%%%%%%%%%%%%%%%%%%%%%%%%%%%%%%%%%%%%%%%%%%%
\subsection{Basic Idea}
%%%%%%%%%%%%%%%%%%%%%%%%%%%%%%%%%%%%%%%%%%%%%

Before going into quivers and mutations,
let us first briefly summarize the basic
idea behind our algorithm of identifying a 3d $\scN=2$
theory from a given 3-manifold, following closely the approach 
of \cite{Terashima:2011qi}. 

\begin{figure}[htbp]
\centering{\includegraphics[scale=0.25]{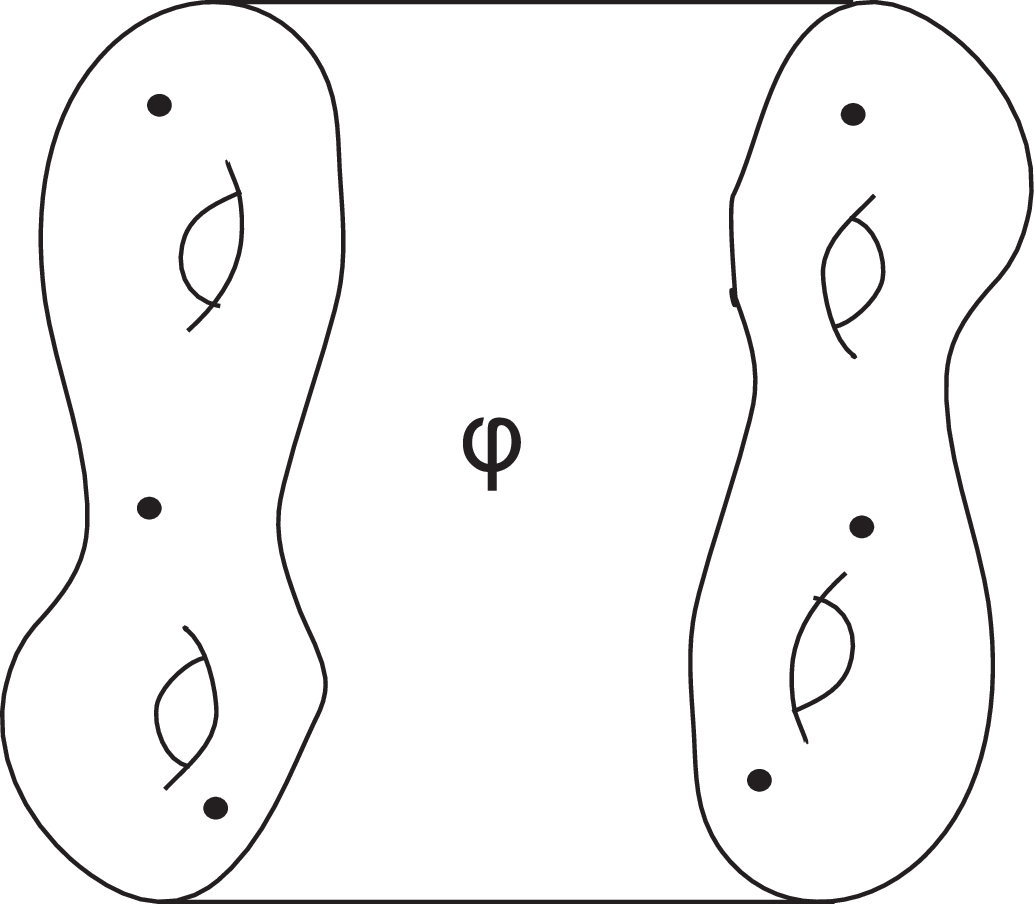}}
\caption{
Canonical quantization of the $SL(2)$ Chern--Simons theory
on $\Sigma_{g,h}\times I$
 gives our
Hilbert space $\scH_{\Sigma}$. In this example $\Sigma$ is a genus $2$
surface with $3$ punctures.
}
\label{fig.timeevolution}
\end{figure}

\begin{figure}[htbp]
\centering{\includegraphics[scale=0.25]{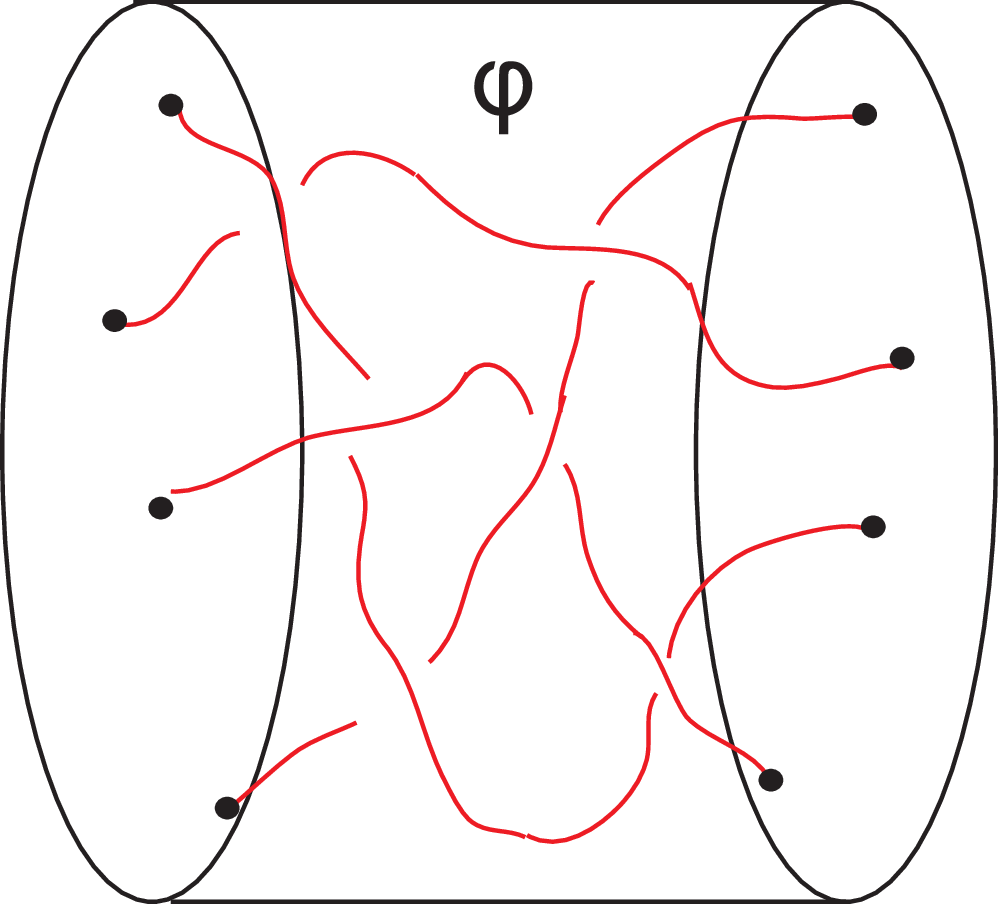}}
\caption{The motion of the punctures represents braids inside the
 mapping cylinder.}
\label{fig.braid}
\end{figure}

Let us first consider a 3-manifold of the form
$\Sigma\times I$, where $\Sigma:=\Sigma_{g,h}$
is a Riemann surface with $h$ punctures and genus $g$,
and $I$ is an interval of finite size (Figure \ref{fig.timeevolution}).
Such a 3-manifold is called a mapping cylinder. 
We will later generalize our analysis to more general 3-manifolds.
We assume
$\Sigma$ is hyperbolic, i.e., $\chi(\Sigma)<0$.
When the surface $\Sigma$ has punctures, the trajectory of the
punctures sweeps out a 1d defect inside the 3-manifold, 
defining braids inside $M$ (Figure \ref{fig.braid}).

We consider $SL(2)$ Chern--Simons theory\footnote{We discuss the holomorphic part of
the $SL(2)$ Chern--Simons theory, and for this purpose it does not matter
whether the gauge group is $SL(2, \bC)$ or $SL(2, \bR)$.} on this 3-manifold (Figure
\ref{fig.timeevolution}).

Since there is a canonical preferred direction on this 3-manifold, we
could regard that direction as a time direction and carry out canonical
quantization, obtaining the Hilbert space of our theory $\scH_{\Sigma}$
on our Riemann surface $\Sigma$. This is identified with the Hilbert
space of the quantum \Teichmuller theory, formulated first in
\cite{Chekhov:1999tn,KashaevQuantization}.\footnote{We do not 
give a detailed explanation for this reason
here--- see the review material in \cite{Terashima:2011qi}. 
Here it suffices to point out that the classical saddle points of
3d $SL(2, \bR)$ Chern--Simons theory are given by flat $SL(2, \bR)$
connections on the 3-manifold, whereas \Teichmuller space is 
a connected
component of moduli space of flat $SL(2, \bR)$ (or rather $PSL(2, \bR)$)
connections on
$\Sigma$, and hence the latter naturally arises in the canonical
quantization of the former.}

Since Chern--Simons theory is topological, the time
evolution is trivial, and the only non-trivial information in this case
is the choice of the boundary conditions. In other words,
corresponding to the two boundaries we need to specify 
the two states
\begin{equation}
|\textrm{in} \rangle , \quad |\textrm{out}\rangle \in \scH_{\Sigma} \ ,
\end{equation}
and the partition function is simply given by the overlap of the two states
\begin{equation}
Z_{\Sigma\times I}=\langle \textrm{in} | \textrm{out} \rangle \ . \label{overlap}
\end{equation}

The coordinate description of $\scH_{\Sigma}$
depends on a choice of an ideal triangulation,
and the in and out states in \eqref{overlap}
might be naturally described in different triangulations.
To take this into consideration we introduce 
an operator $\hat{\varphi}$ representing the change of the triangulation
between in and out states, leading to\footnote{
We can describe this more formally. Given two triangulations $T, T'$,
there is an operator $\hat{\varphi}_{T',T}$ such that
\begin{equation*}
|\textrm{out}\rangle_{T'}=\hat{\varphi}_{T', T}\,
|\textrm{out}\rangle_{T} \ ,
\label{varphiaction}
\end{equation*}
and the partition function is given by
\begin{equation*}
Z_{\Sigma\times I}=\langle \textrm{in} |_T  |\textrm{out}\rangle_{T'}
=\langle \textrm{in}|_T\,  \left(\hat{\varphi}_{T',T}\right) \,  
|  \textrm{out} \rangle_T \ .
\end{equation*}
The difference between $| \, \rangle_{T}$ and $| \, \rangle_{T'}$ will be crucial when 
we identify the boundary data, see \eqref{mappingtorus} and section \ref{sec.cap}.
}
\begin{equation}
Z_{\Sigma\times I}
 =\langle \textrm{in}|\,  \hat{\varphi}
 \,  |  \textrm{out} \rangle \ .
\label{inphiout}
\end{equation}
Geometrically $\varphi$ will be an element of the mapping class group 
of $\Sigma$, i.e.\ $\varphi$ is a large coordinate transformation on $\Sigma$.

We can also identify the in and out states, and take a sum over all the
possible states. 
\begin{equation}
Z_{(\Sigma\times S^1)_{\varphi}}
 =\textrm{Tr}(\varphi)= \int \! d(\textrm{in})\, \langle \textrm{in} | \hat{\varphi}
 \,  |  \textrm{in} \rangle \ .
\label{mappingtorus}
\end{equation}
In this case the geometry is that of a mapping torus $(\Sigma\times
S^1)_{\varphi}$, defined by identifying $(x, 0)\sim (\varphi(x), 1)$
for $\Sigma\times I$. Here we have taken $I=[0,1]$, and $\varphi$
to be an element of the mapping class group of $\Sigma$.

Since the action of $\hat{\varphi}$ is given explicitly, we can evaluate this
expectation value and obtain an integral expression --- 
the integrand contains one quantum dilogarithm function
\cite{FaddeevVolkovAbelian,FaddeevKashaevQuantum,Faddeev95}
for each 
flip of the triangulation.
We can then read off the corresponding 3d $\scN=2$ 
gauge theory, using the relation \eqref{ZZ} as a guideline.
The resulting 3d $\scN=2$ theory can
be thought of as a theory on the duality domain wall theory \cite{Gaiotto:2008sa,Gaiotto:2008ak}
inside the 4d $\scN=2$ theory of \cite{Gaiotto:2009we}.

While the strategy outlined to this point should
work,
this program has never been worked out in generality.
It is also the case that the resulting 3-manifold is apparently limited to 
 mapping cylinders or mapping tori,
and it is not clear if this method generalizes to more
 general 3-manifolds.

The goal of this section is to fill in these gaps.

%%%%%%%%%%%%%%%%%%%%%%%%%%%%%%%%%%%%%%%%%%%%%
\subsection{Quantum \Teichmuller Theory}\label{sec.Teichmuller}
%%%%%%%%%%%%%%%%%%%%%%%%%%%%%%%%%%%%%%%%%%%%%

The construction of the Hilbert space $\scH_{\Sigma}$ relies on the 
quantum \Teichmuller theory,
which fits neatly into the general framework
of the previous sections \cite{GSV,FST1}.

\begin{figure}[htbp]
\centering{\includegraphics[scale=0.35]{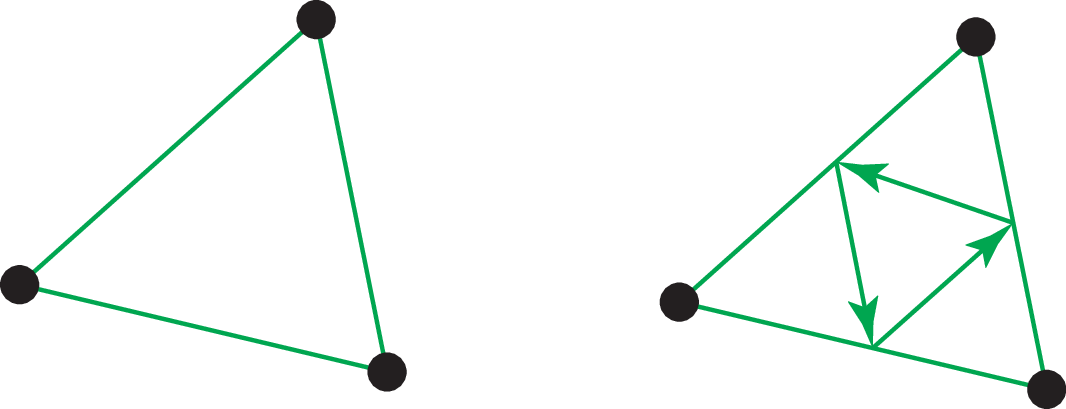}}
\caption{
Given a triangulation of a 2d surface, we can associate a quiver by
 drawing the quiver in this figure for each triangle.
A vertex of the quiver is associated with an edge of the triangulation.
}
\label{nee}
\end{figure}

\begin{figure}[htbp]
\centering\includegraphics[scale=0.38]{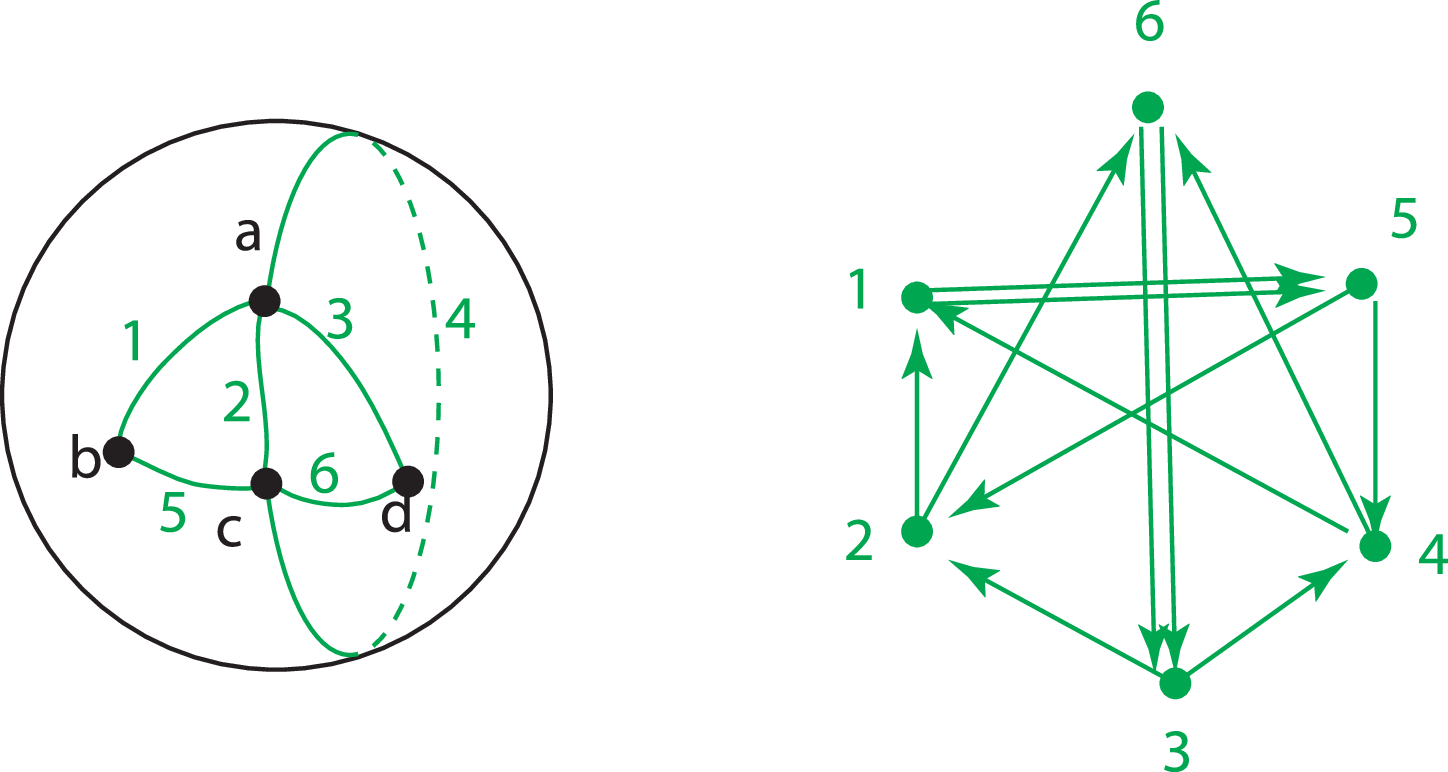}
\caption{An example of a quiver defined for a triangulation, for a
 $5$-punctured sphere. The quiver is obtained by 
 applying the rule of Figure \ref{nee} for each triangle.}
\label{fig.quivertrig}
\end{figure}

Suppose that we have a punctured Riemann surface with negative Euler character.
Let us choose an ideal triangulation $T$ of the surface, 
i.e., a triangulation such that all the vertices are at the punctures.
Given a triangulation of a 2d surface, we can associate a quiver by
 drawing a 3-node quiver for each triangle (Figure \ref{nee}).
The index set of this quiver $i,j, \ldots \in I$
is the set of edges,
and the matrix $Q$ satisfies
\begin{align}
Q_{i,j}\in \{-2, -1,0,1,2\} \ ,
\end{align}
for all $i,j\in I$.
See Figure \ref{fig.quivertrig} for an example.

Given this quiver we could construct the algebra $\scA_Q$ and the
Hilbert space $\scH_Q$ as in the previous
subsection. 
In the \Teichmuller language, $\sfx_i$ is the quantization of the Fock
(shear) coordinate \cite{Fock}, which is the coordinate of the 
\Teichmuller space. The commutation relation \eqref{xCCR}
represents the standard symplectic form (Weil--Petersson form)
on the \Teichmuller space,
and the Hilbert space $\scH_Q$
coincides with the Hilbert space of the 
quantum \Teichmuller theory.

\begin{figure}[htbp]
\centering\includegraphics[scale=0.38]{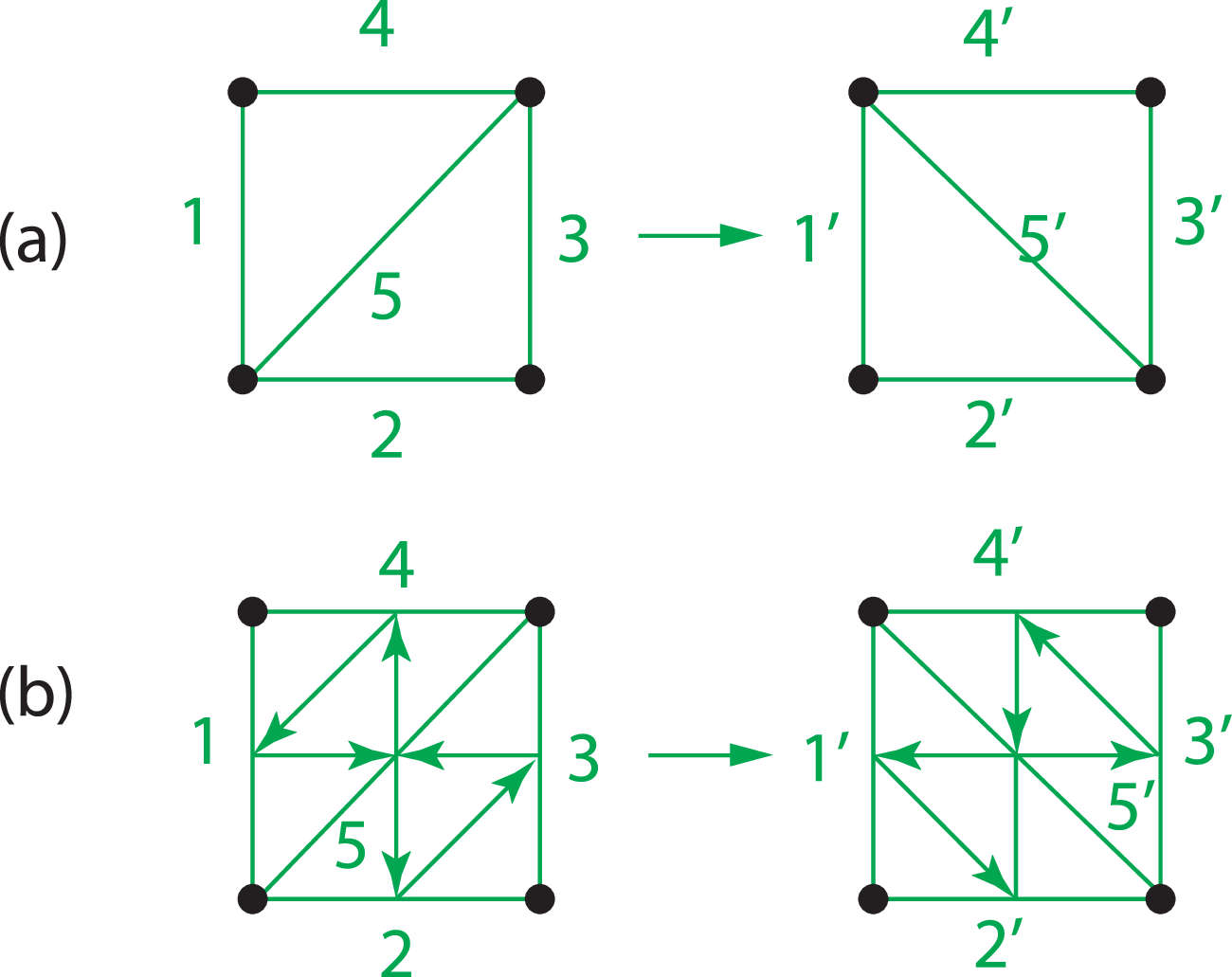}
\caption{(a) A flip changes a diagonal of a square, consisting of two
 triangles in the triangulation. (b) A flip corresponds to a mutation of
 the associated quiver.}
\label{fig.flip}
\end{figure}

The description to this point relies on the choice of a 
triangulation. We can change the triangulation by a flip,
a change of a diagonal of a square (Figure \ref{fig.flip}).
In fact, it is known that any two ideal triangulations are 
related by a sequence of flips.

It is easy to see that the effect of such a flip is translated into a
mutation of the associated quiver diagram.
In this case, the mutation rule \eqref{ymutation}
simplifies to
\begin{equation}
\begin{split}
&\sfx_{1}\to  \sfx_1 (1+q\, \sfx_5)  \ ,\quad 
\sfx_{2}\to \sfx_2 (1+q^{-1}\sfx_5^{-1})^{-1} \ , \\
&\sfx_{3}\to  \sfx_3 (1+q\, \sfx_5) \ , \quad
\sfx_{4}\to \sfx_4 (1+q^{-1}\sfx_5^{-1})^{-1}  \ , \quad
\quad \sfx_5\to \sfx_5^{-1} \ ,
\end{split}
\end{equation}
where we use the labeling in Figure \ref{fig.flip}.

To make contact with the discussion of the previous subsection,
note that an element of the mapping class group $\varphi$ 
changes the triangulation,
%on $\scH_{\Sigma}$.
%This changes the triangulation,
and this in turn is represented by a
sequence of flips $\bm{m}=(m_1, \ldots, m_L)$. We could then define the 
associated partition function $Z^{\rm cluster}_{(\Sigma, \varphi)}$ to
be $Z^{\rm cluster}_{(Q,
\bm{m})}$ defined in \eqref{Zeb}.

Note that given $\varphi$ the choice of flips $\bm{m}$ is far from
unique.
However, different choices of $\bm{m}$ for a given $\varphi$ lead to
the same partition function, thanks to the 
quantum dilogarithm identities \cite{FockGoncharovQuantumCluster,KashaevNakanishi}.\footnote{
We need to keep track of the labeling of edges in order to write down quantum dilogarithm identities.
}

%%%%%%%%%%%%%%%%%%%%%%%%%%%%%%%%%%%%%%%%%%%%%
\subsection{Canonical Ideal Triangulations}\label{sec.hyperbolic}
%%%%%%%%%%%%%%%%%%%%%%%%%%%%%%%%%%%%%%%%%%%%%

Let us discuss the ideal triangulation of $M$.
For a mapping cylinder there is a canonical choice of ideal 
triangulation \cite{FloydHatcher,Lackenby,Gueritaud}.

\begin{figure}[htbp]
\centering{\includegraphics[scale=0.29]{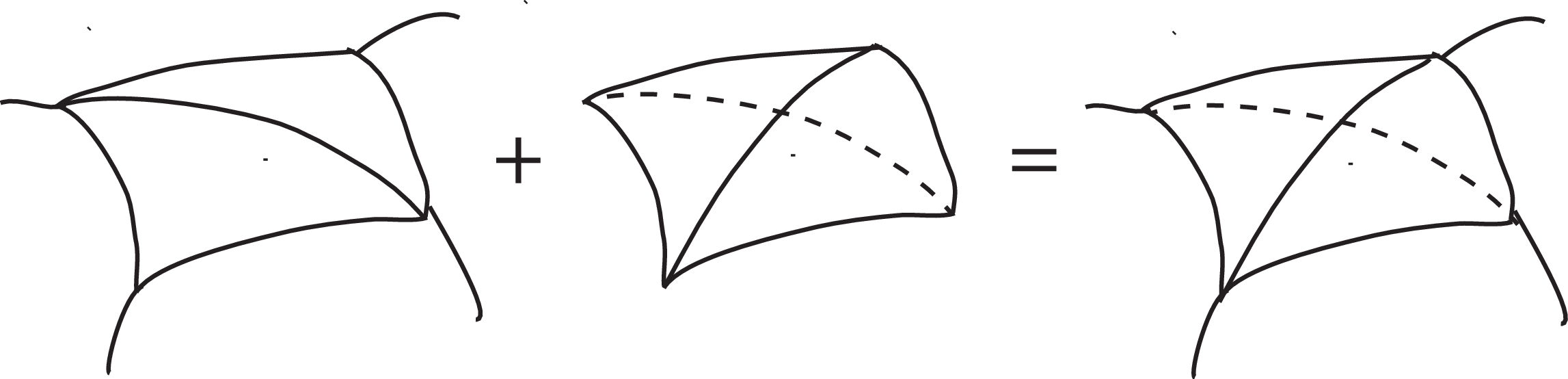}}
\caption{Attaching a 3d tetrahedron to a 2d triangulation
flips the diagonal of a square.}
\label{fig.pillow}
\end{figure}

As we have seen already, the action of $\varphi$
could be traded for a sequence of flips (which in turn is identified
with a mutation sequence $\bm{m}$).
We can then associate a tetrahedron 
for each flip; given a 2-manifold with triangulation, 
we can attach a tetrahedron (squeezed like a pillowcase)
and we effectively obtain a new 2-manifold with a different
triangulation, 
related to the original one by a flip (Figure \ref{fig.pillow}). 
By repeating this procedure we obtain 
a sequence of tetrahedra, whose faces are glued together.
The 3-manifold is now decomposed into tetrahedra:
\begin{align}
M=\bigcup_m \Delta_m \ .
\end{align}

Our mutation network contains
all the information about canonical triangulations.
We associate an ideal tetrahedron for each mutation $m$ inside $\bm{m}$
(represented by a black vertex).
Since a tetrahedron has six edges, each black vertex is connected with 
six white vertices (see the right figure of Figure \ref{aroundwhite}). 
The mutation network also specifies how to glue these tetrahedra
together, and hence the gluing equations in Appendix \ref{sec.brief};
an edge $w$ (a white vertex) is 
shared by all the tetrahedra (black vertices)
which are connected with the 
$w$ in the mutation network (see also Figure
\ref{fig.gluetet} in section \ref{sec.structure}).

The canonical triangulation is so far a combinatorial triangulation, however,
we can promote it to an ideal triangulation
by the hyperbolic tetrahedron
when the 3-manifold is hyperbolic.\footnote{
A 3-manifold is ``generically'' hyperbolic; a knot complement 
in $S^3$, to be discussed in the next subsection, is hyperbolic unless the knot is one of the torus knots
or their  satellites.
}
This means that each tetrahedron is an ideal tetrahedron in $\bH^3$,
and there is a (complete) hyperbolic structure of the 
3-manifold (see Appendix \ref{sec.brief} for brief summary of 
the 3d hyperbolic geometry needed for this paper).

\begin{figure}[htbp]
\centering{\includegraphics[scale=0.33]{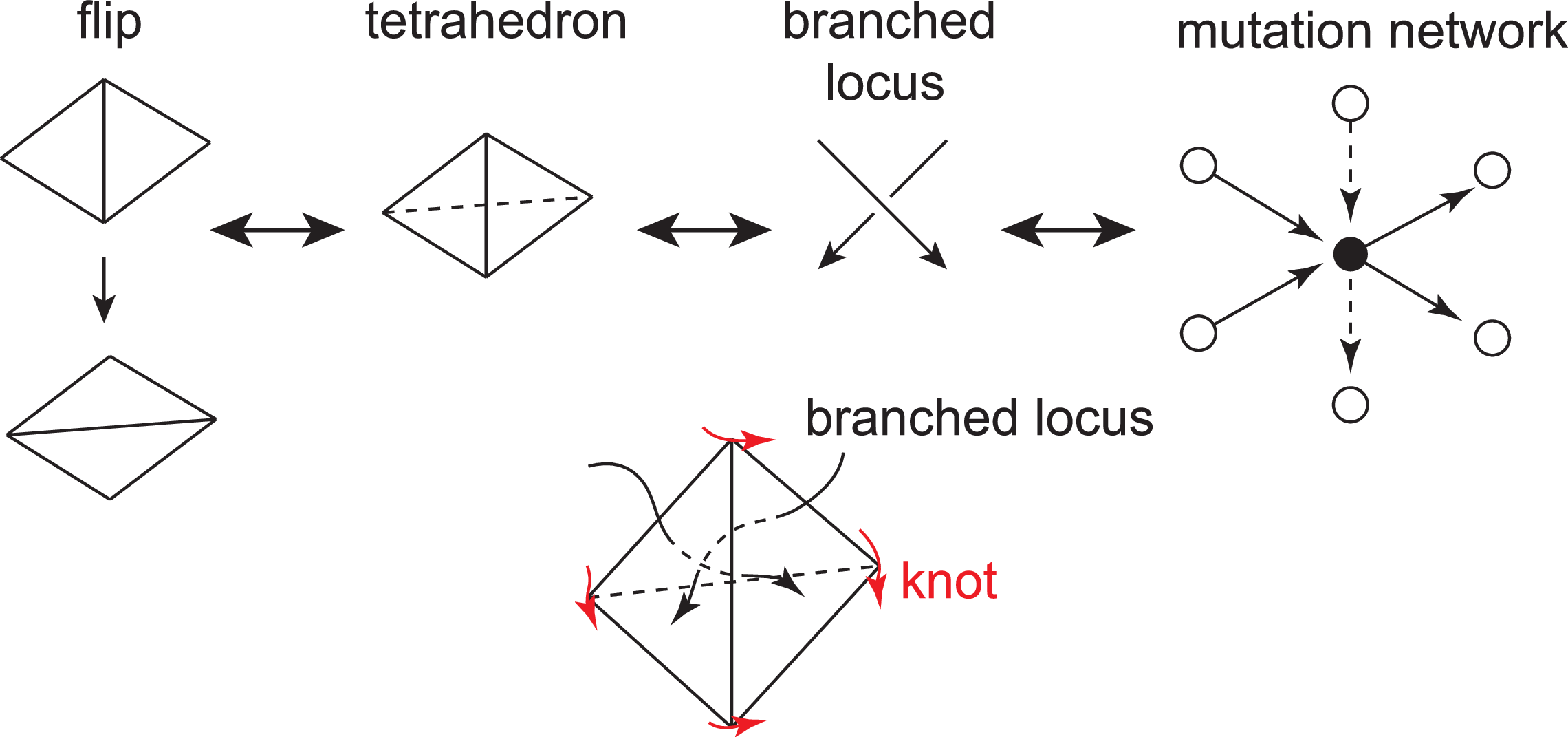}}
\caption{A flip corresponds to a tetrahedron, which is represented as a 
crossing in the branched locus of \cite{Cecotti:2011iy}, or a black
 vertex of the mutation network introduced in this paper.
 Beware of the difference between the branched locus of \cite{Cecotti:2011iy} and 
 the braids in this paper.}
\label{fig.basic}
\end{figure}

The mutation network in the 3-manifold cases discussed in this section
is reminiscent of the ``braid/tangle''
of \cite{Cecotti:2011iy}. This is the branched locus of the 
IR geometry, which is a double cover of our 3-manifold $M$.
In both cases we associate a basic building block, either a black vertex
(for mutation network) or a crossing, to a tetrahedron (Figure \ref{fig.basic}).

However, it is important to keep in mind that 
the ``braid'' in their paper, or rather the branched locus, is
\emph{not}
the braid/knot discussed in our paper. In fact, inside an
ideal hyperbolic tetrahedron our knot (which appear in ``knot
complement''), for example, goes through the vertices of tetrahedra,
whereas the ``braid'' in \cite{Cecotti:2011iy} goes though
the faces of tetrahedra (Figure \ref{fig.basic}).
In other words their ``braid'' pass through the zeros of quadratic differential
of a 2d surface in the section of the 3-manifold,
whereas our braids pass through the poles.
In the rest of this paper the words ``knot/link/braid''  will always
 refer to the 
knot/link/braid in our sense.

%%%%%%%%%%%%%%%%%%%%%%%%%%%%%%%%%%%%%%%%%%%%%
\subsection{Capping the Braids}\label{sec.cap}
%%%%%%%%%%%%%%%%%%%%%%%%%%%%%%%%%%%%%%%%%%%%%

There is a caveat in the discussion to this point.
The 3-manifold obtained in this construction is of special type
$\Sigma\times I$, and does not seem to be general enough.
However, what saves the day is that 
by suitably identifying boundaries of this 3-manifold
it is possible to obtain a rather large class of 3-manifolds,
including all the link complements in $S^3$.

What we do here is to identify unglued faces of tetrahedra
on the boundary of the mapping cylinder.
Depending on the identification
we obtain different 3-manifolds.

\begin{figure}[htbp]
\centering{\includegraphics[scale=0.4]{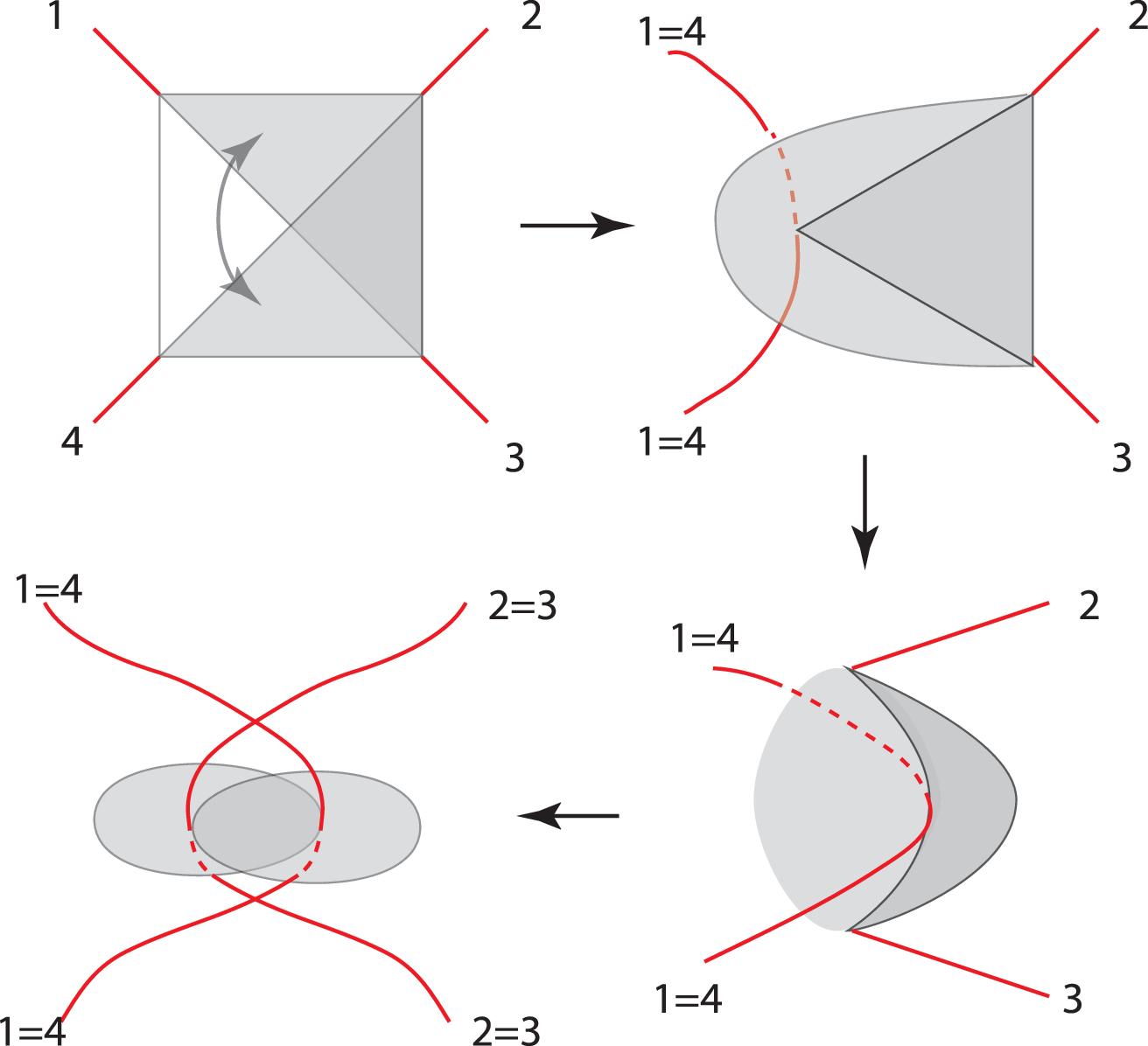}}
\caption{By identifying the faces of $4$-punctured sphere bundles, 
we can identify the four braids (corresponding to four punctures) in
 pairs; the same figures can be found in \cite{SakumaWeeks,FuterGueritaud}.}
\label{fig.cap}
\end{figure}

Such an identification has been worked out for the case of the 
$4$-punctured sphere \cite{SakumaWeeks,FuterGueritaud}. 
In this case, there are four faces in the triangulation, and 
we first identify two of the faces and then the remaining two
(see Figure \ref{fig.cap}). We can verify that this face identification 
gives rise to the identification of the braids passing through
the four punctures, and hence the braids are capped off into
links.
We can realize a large class of knots called 2-bridge knots in this way.

\begin{figure}[htbp]
\centering{\includegraphics[scale=0.4]{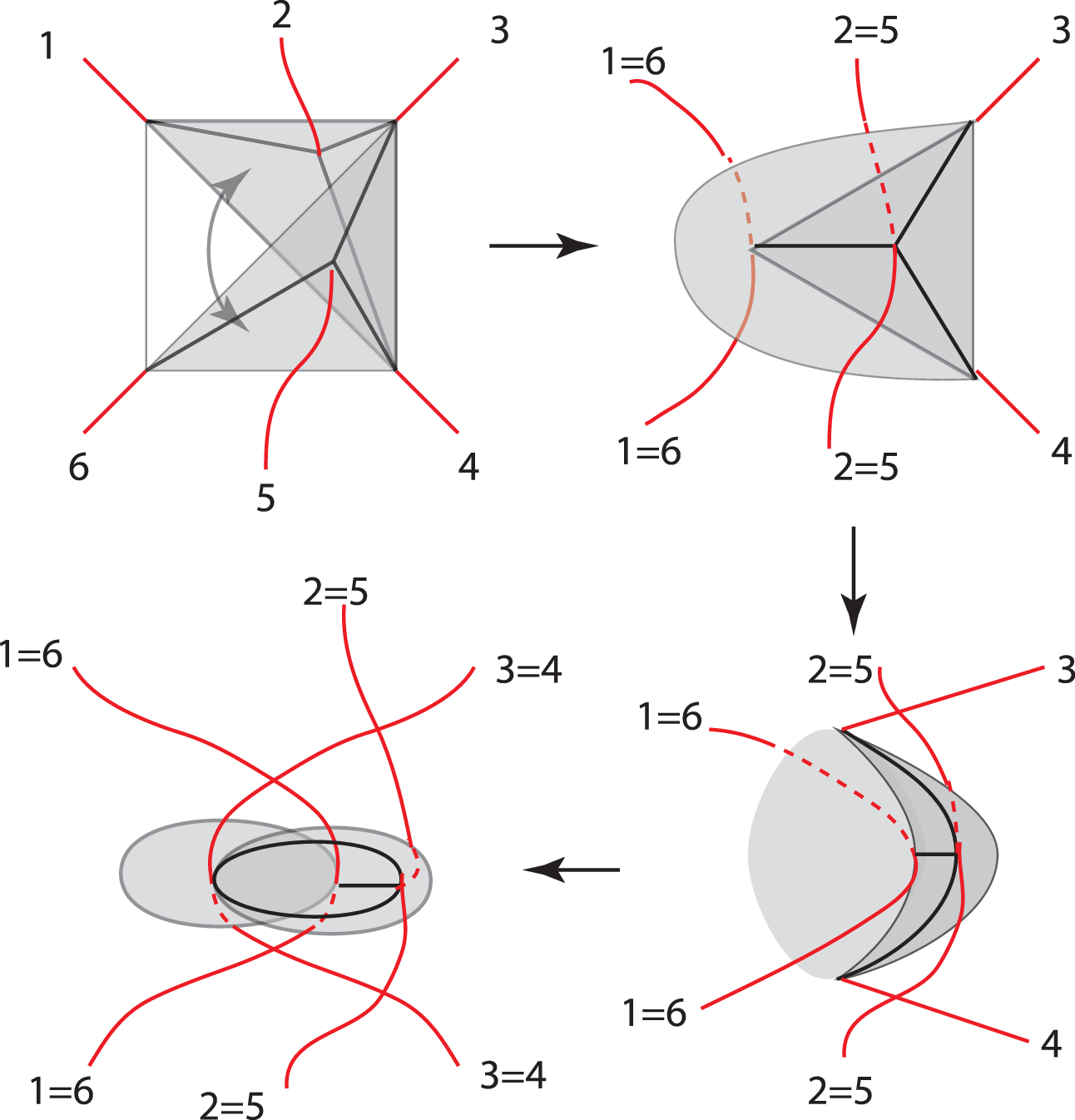}}
\caption{Similar identification exists for $2n$-punctured sphere
 bundles, 
by iteratively making a pair among the faces. This is an example of $n=3$.}
\label{fig.cap2}
\end{figure}

We can generalize this construction to $2n$-punctured spheres (Figure \ref{fig.cap2}).
In this case we can again identify the faces of the boundary surface,
leading to the identification of the braids.
By applying an element of the mapping class group we 
could obtain an arbitrary identification of $2n$-braids.

\begin{figure}[htbp]
\centering{\includegraphics[scale=0.29]{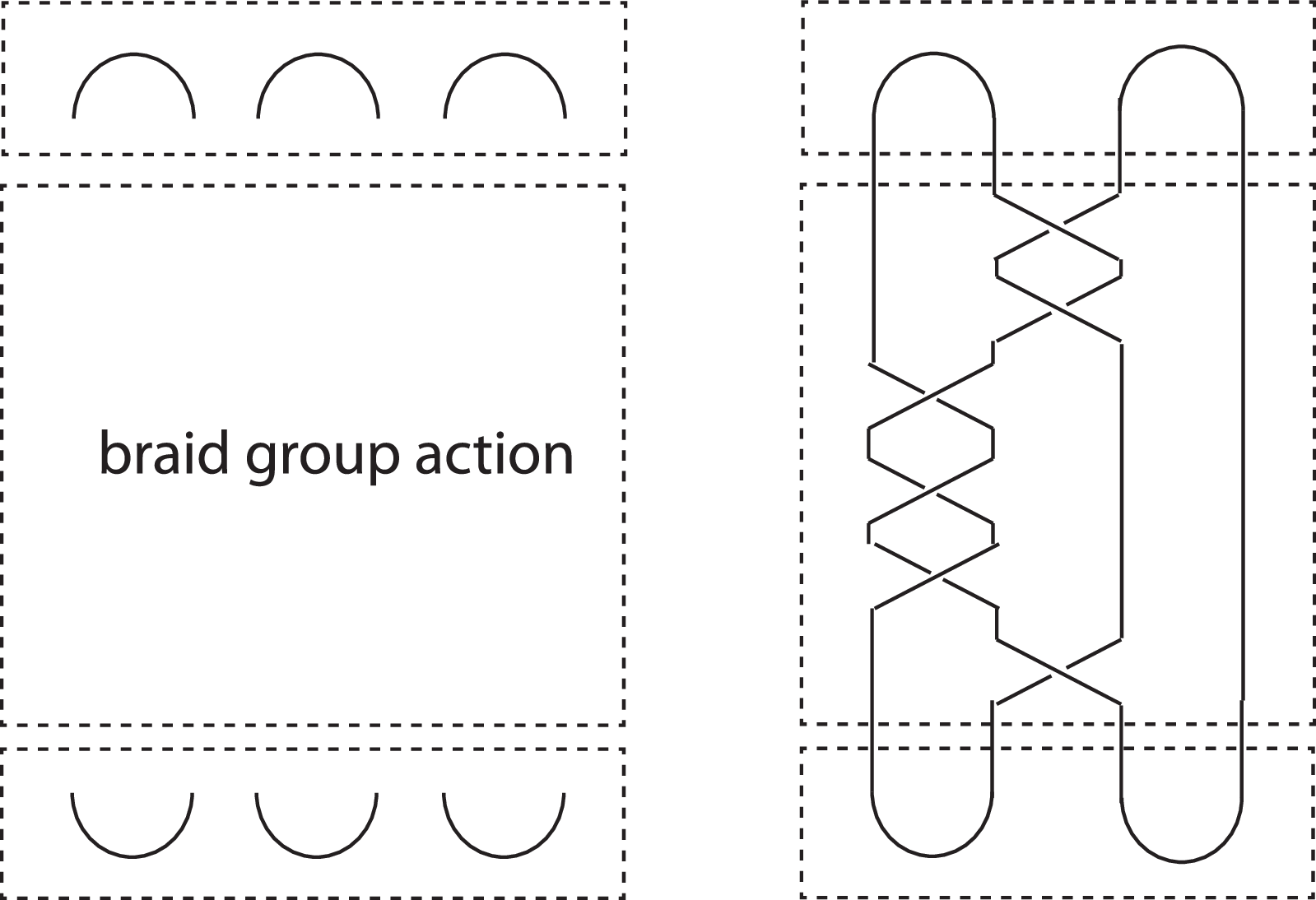}}
\caption{A $2n$-plat representation of a knot/link. The link is
 determined by an element of the the braid group $\scB_{2n}$.
The example on the right is determined by an element $s_1^2 s_2^3
 s_1^2$ of $\scB_6$, and is a link often called $6_1$.}
\label{fig.plat}
\end{figure}

In particular we could choose the identification as in Figure
\ref{fig.plat}
for the $2n$-punctured sphere bundle, both for the 
in and out states.
A link obtained after such an identification
is said to have $2n$-plat representation.\footnote{
This is similar to, but different from, the so-called braid representation of a knot/link.
}
It is known that an arbitrary link has a $2n$-plat representation for
some $n$ \cite{Birman}, which means that our procedure includes an arbitrary
link in $S^3$.
It is clear from Figure \ref{fig.plat}
that the resulting link is determined from
an element of the braid group $\scB_{2n}$ 
(recall also Figure \ref{fig.braid}).

The identification of the faces 
induces identification of edges,
which is translated into the identification of 
corresponding variables $\sfx_i$ in $\scA_Q$ and $\scH_Q$.\footnote{
Strictly speaking this identification could involve
factors of $q$ as long as they become trivial in the 
classical limit. These factors affect 
our partition function, but not 
the semiclassical analysis of section \ref{sec.structure}.
}
In mutation networks this procedure of gluing boundary faces of mapping
cylinders is simply translated into the identifications of 
white vertices (edges of tetrahedra).
We will discuss the example of the figure-eight knot complement in section \ref{sec.example}.

\subsubsection*{Comparison with Heegaard Decomposition\footnote{
This part is outside the main track of this section and could be skipped on first reading.
}
}
Our capping procedure is closely related with the
Heegaard decomposition of a 3-manifold, and its generalization.

\begin{figure}[htbp]
\centering{\includegraphics[scale=0.3]{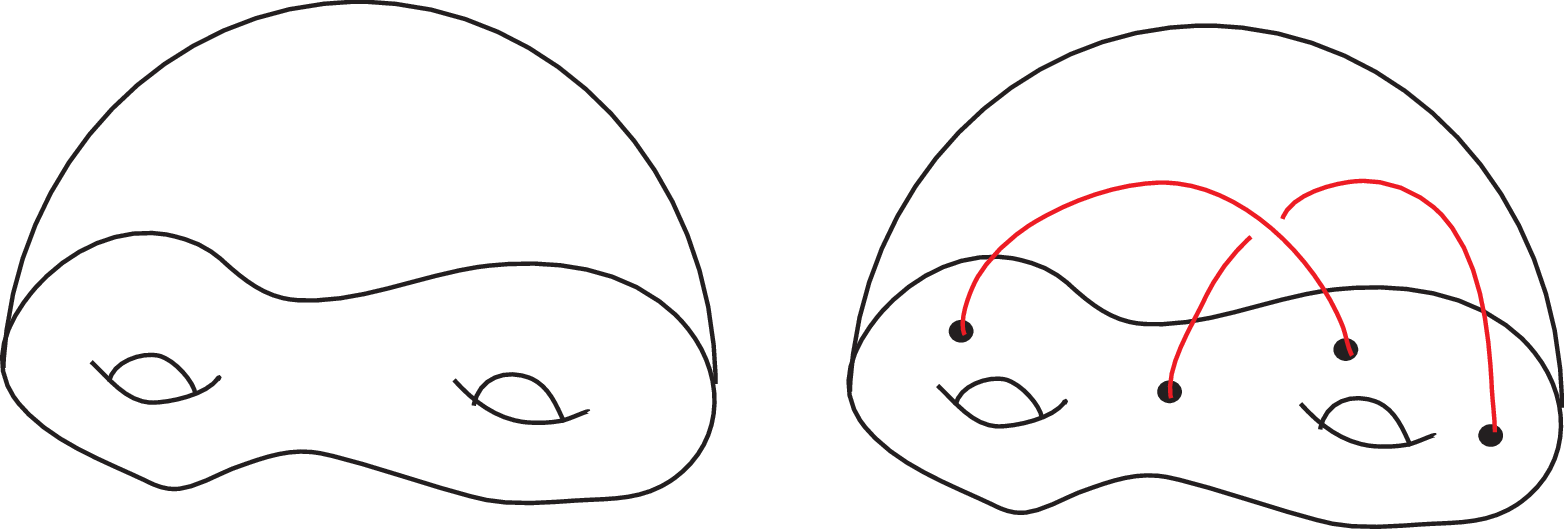}}
\caption{A handlebody (left) and a tanglebody (Right).}
\label{fig.tanglebody}
\end{figure}

A Heegaard decomposition states that a closed 3-manifold $M$
has a decomposition of the form
\begin{align}
M=H_1 \cup_{\varphi} H_2\ , \quad
\partial H_1=\partial H_2=\Sigma_{g,0} \ , 
\label{Heegaard}
\end{align}
where $H_1$ and $H_2$ are the handlebodies\footnote{Colloquially 
they are the ``simplest'' 3-manifolds with boundary 
$\Sigma$ (the left of Figure \ref{fig.tanglebody}). For example, 
the handlebody for the two-sphere $S^2$
is the three-dimensional ball $B^3$.}
and $\varphi$ is an element of the mapping class group of $\Sigma_{g,0}$.

\begin{figure}[htbp]
\centering{\includegraphics[scale=0.23]{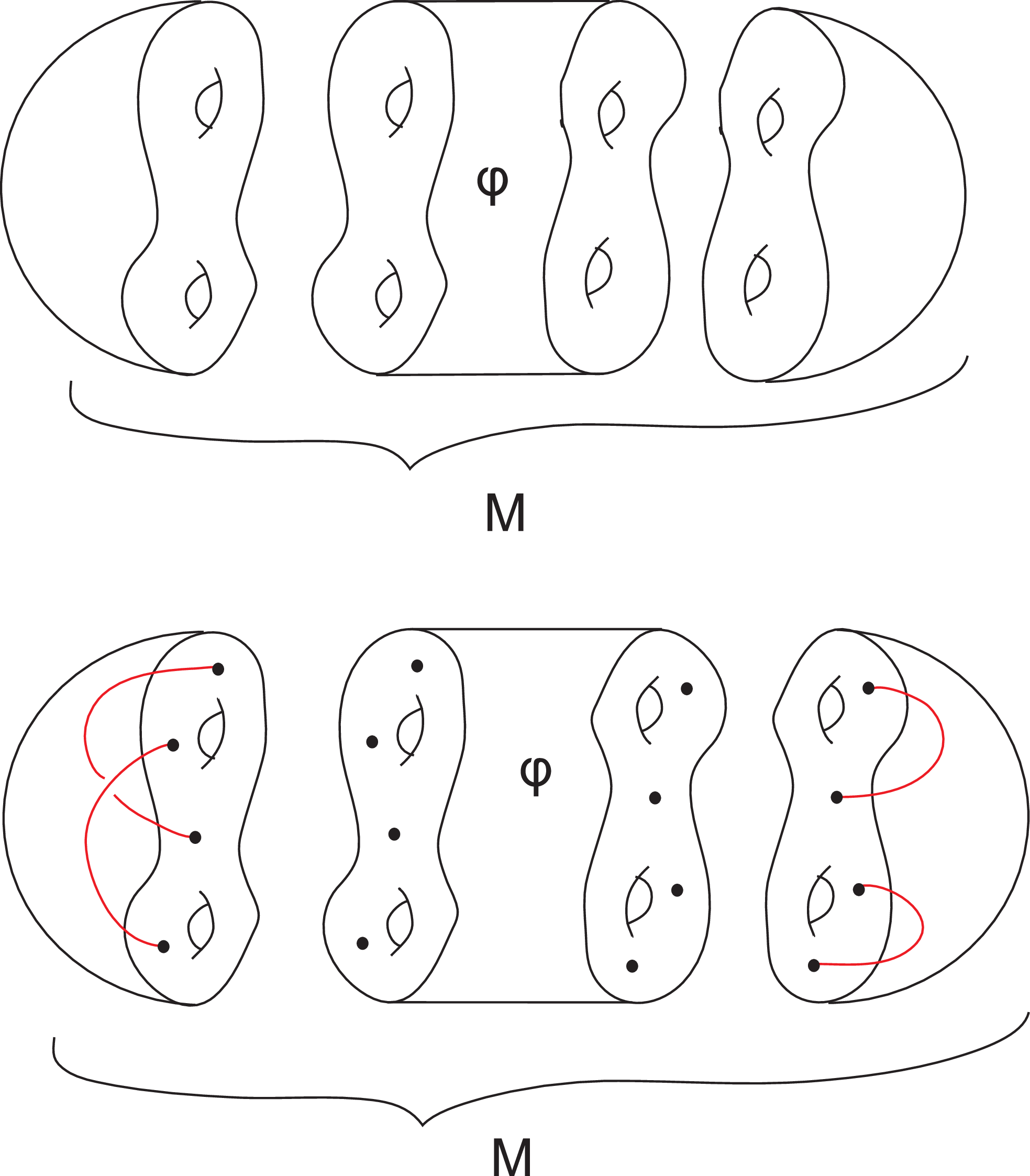}}
\caption{Heegaard decomposition (a) and tanglebody decomposition (b).}
\label{fig.Heegaard}
\end{figure}

The decomposition \eqref{Heegaard} could be represented as in Figure
\ref{fig.Heegaard} (a).
As the figure shows, the only effect of the handlebody should be to choose a 
specific element 
\begin{equation*}
|\textrm{handlebody} \rangle \in \scH_{\Sigma}  \ ,
\end{equation*}
and we could evaluate the partition function by
substituting these states in the in and out states \eqref{inphiout}.

For our purposes we still need a small modification; we need to include
knots, and consider a 3-manifold with torus boundaries.
In the two-dimensional slice the knots are point-like in 
the two-dimensional surface $\Sigma$, and serve as a puncture of $\Sigma$ 
(recall Figure \ref{fig.braid}).

\begin{figure}[htbp]
\centering{\includegraphics[scale=0.35]{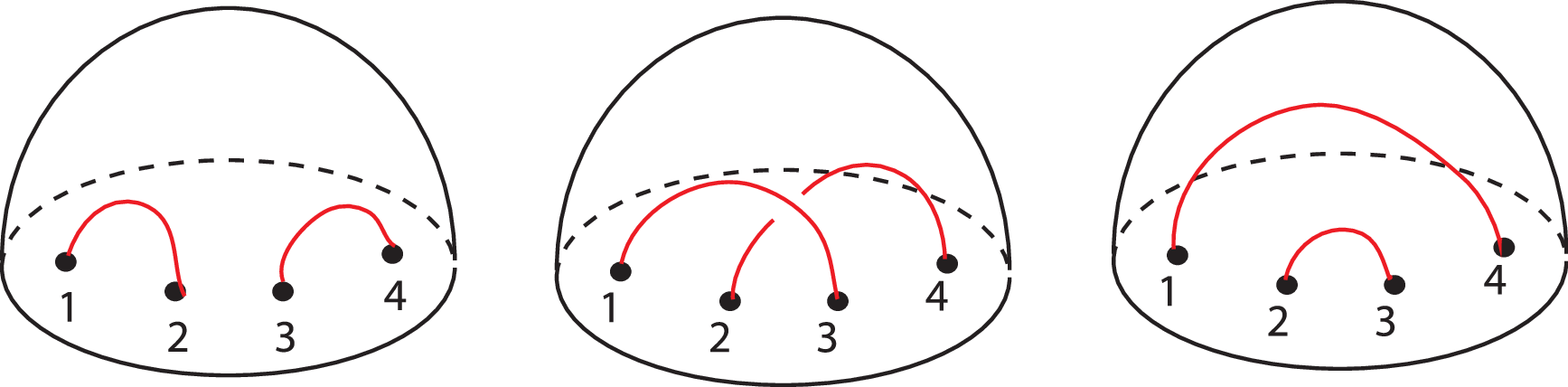}}
\caption{Three tanglebodies for the 4-punctured sphere.}
\label{fig.skein}
\end{figure}

Correspondingly, we need to consider a handlebody with knots inside. We
call these a tanglebody: see right of Figure \ref{fig.tanglebody}.\footnote{
As commented already, a handlebody for a sphere is simply a ball, and in
this case it is straightforward to define the corresponding tanglebody
as a ball with braids deleted from it.
}
Note that the tanglebody exists only when the number of punctures of
$\Sigma$ is even, since whenever a knot comes into the tanglebody it
needs to come out. It is also clear that given $\Sigma$
the tanglebody is not unique. For example, in the case of the 
$4$-punctured sphere we have the three tanglebodies of 
Figure \ref{fig.skein}, and each of them gives rise to different states.
Note that the corresponding choice is present in Figure \ref{fig.cap},
where we identify the four faces of $4$-punctured spheres in pairs.

Now we can generalize the Heegaard decomposition and 
consider the ``tanglebody decomposition'' (Figure \ref{fig.Heegaard})
\begin{align}
M=T_1 \cup_{\varphi} T_2\ , \quad
\partial T_1=\partial T_2=\Sigma_{g,h} \ , 
\label{tangledecomp}
\end{align}
where $T_{1,2}$ are tanglebodies.

Our gluing procedure explained above (Figures \ref{fig.cap} and \ref{fig.cap2})
should be essentially the same as 
gluing the tanglebodies, in the sense that in both cases 
the braids are identified in the same way. 
The generality of the Heegaard decomposition
roughly explains the generality of our approach. 
It would be interesting, however, to understand the relation between the
two approaches in more detail.

%%%%%%%%%%%%%%%%%%%%%%%%%%%%%%%%%%%%%%%%%%%%%
\subsection{The Semiclassical Limit}\label{sec.structure}
%%%%%%%%%%%%%%%%%%%%%%%%%%%%%%%%%%%%%%%%%%%%%

Finally, let us directly verify that the partition function discussed in the
previous section reproduces the gluing equations \eqref{structure}
of the associated hyperbolic $3$-manifold.
This is a direct demonstration of the consistency between
this section and section \ref{sec.cluster}.
The semiclassical analysis of our partition functions can be found in 
\cite{KashaevNakanishi},\footnote{
Our semiclassical analysis here is actually slightly 
different from \cite{KashaevNakanishi}
in that we have used \eqref{Zresult} with the variables $p_i(t)$ integrated out,
whereas they used \eqref{KNborrow} and extremized also with respect to $p_i(t)$.
Both methods give essentially the same results. 
In fact, the proof of \eqref{sumZ} here is somewhat similar
the proof in \cite{Nagao:2011aa}, although proven in different variables.
}
and the fact that the saddle point analysis of our partition function 
reproduces the gluing equations, as well as the connection with 
cluster algebras, has already been worked out in \cite{Nagao:2011aa};
see also \cite{HikamiInoue}.

The classical limit of the Chern--Simons theory is $t\to \infty$,
or equivalently the $b\to 0, q\to 1$ limit of \eqref{ZZ}.
It is straightforward to take the $b\to 0$ limit of the expression \eqref{Zmain}
obtained in the previous section, and 
we find, using \eqref{ebclassical},
\begin{align}
Z_{(\Sigma, \varphi)} \to \int \prod_{w\in W_{\rm int}} dx_w \,\,\exp\left( \frac{\i}{2\pi b^2} \scW(x_w) \right)\ ,
\end{align}
where $\scW(x_w)$ is written as a sum over the contributions $\scW^{(m)}(x_w)$, each associated
with a flip $m$:
\begin{align}
\scW(x_w)&=\sum_{m\in B} \scW^{(m)}(x_w) \nonumber \\
&=\sum_{m\in B} \left(\Li_2(e^{Z'(m)})-\half
 Z'(m)Z''(m) 
 \right) \ .
\label{Wcontribution}
\end{align}
The saddle point of this integral is given by
\begin{align}
\exp\left( \frac{\partial \scW}{\partial x_w}\right)=1 , \quad w\in
 W_{\rm int}\ .
\label{2dsaddle}
\end{align}
We now claim that this equation is identical to the gluing equation
for the canonical ideal triangulation, and that  
$\scW(x_w)$ is a generating function of the gluing equations
described in \cite{NeumannZagier}.

Let us pick up a particular edge of the triangulation.
This is represented by a vertex $w\in B$.
Then the $x_w$-dependent part of $\scW$
is a sum over contributions (denoted by $\scW^{(m)}$)
from 
all the tetrahedra $m$ containing the edge $e$,
where $\scW^{(m)}$ in this case is given by \eqref{Wcontribution}.

The mutation $m$ is divided into three types, of type 1, type 2 or type 3, as discussed in section \ref{sec.network}.
The $x_w$-derivative of the $\scW^{(m)}$ in each of the three cases 
is given by
\begin{align}
\begin{aligned}
%\begin{split}
&2\log(z^{(m)})-i\pi-Z(m), \\
 %\! \!\quad 
& [-Q_{m,a}]_+ (2\log(z'^{(m)})-Z'(m)), \\
%\!\!\quad 
&[Q_{m,a}]_+ (2\log(z''^{(m)})-Z''(m)) \ ,
\label{derW}
%\end{split}
\end{aligned}
\end{align}
for type 1, type 2, and type 3, respectively,
where we introduced tetrahedron modulus for the tetrahedron $m$
by
\begin{align}
z'^{(m)}:=e^{Z'(m)} \ ,
\end{align}
and
we introduced the three parameters $z, z', z''$
related by
$$
z'^{(m)}=1/(1-z^{(m)}), \quad
z''^{(m)}=1-z^{(m)}{}^{-1} \ .
$$
These will correspond to three different parametrizations of a
tetrahedron,
as explained in Appendix \ref{sec.brief}.

When we collect these factors and sum over $m$, 
the terms linear in $Z(m), Z'(m), Z''(m)$ in \eqref{derW} cancel out due to \eqref{sumZ}
(note $b\scQ\to 1$ in the semiclassical limit).
This means that we are left with
\begin{align}
\exp \left( \fpp{\scW(w)}{w} \right)=\prod_{m:\, {\rm type\ 1}} z^{(m)} 
\prod_{m:\, {\rm type\ 2}} (z^{'(m)})^{[-Q_{m,a}]_+} 
\prod_{m:\, {\rm type\ 3}} (z^{''(m)})^{[Q_{m,a}]_+} \ .
\label{saddlepoint}
\end{align}
When our theory is associated with the 3-manifolds,
this is exactly the gluing equation \eqref{structure}
in section \ref{sec.hyperbolic},
and our derivation represents the fact (proven in \cite{Nagao:2011aa})
that $y$-variables automatically solve the gluing equations.

Interestingly,
there is a natural symmetry cyclically exchanging the 
$Z(m), Z'(m)$ and $Z''(m)$. 
In fact, one can replace \eqref{Wcontribution} by
\begin{align}
\scW'(x_w)&=\sum_{m\in B} \left(\Li_2(e^{Z(m)})-\half
 (Z(m)-\i \pi)Z'(m) 
 \right) \ ,
\label{Wcontribution2}
\end{align}
and we can verify that 
this still gives the correct saddle point equation.
For the 3-manifold cases we can understand this symmetry as the
cyclic exchange of three modulus parametrization $z, z', z''$ (\eqref{ztriple}
in Appendix \ref{sec.brief}).

\begin{figure}[htbp]
\centering\includegraphics{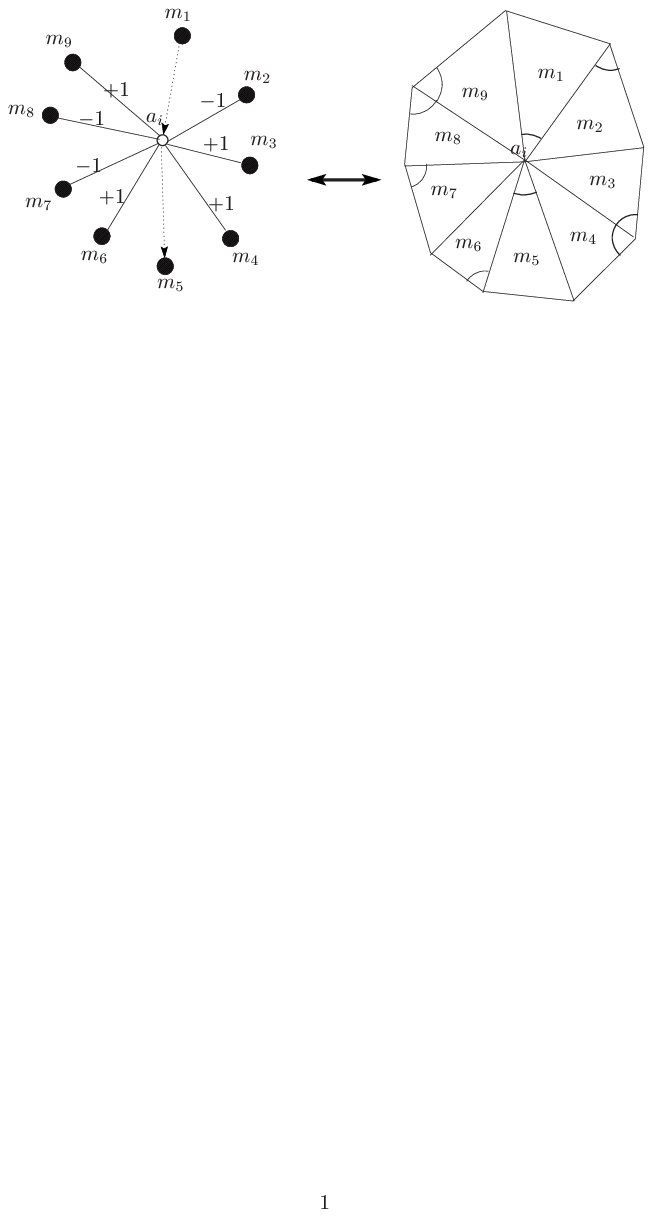}
\caption{The mutation network around a white vertex represents 
how the tetrahedra are glued together.}
\label{fig.gluetet}
\end{figure}

\begin{figure}[htbp]
  \centering
  \includegraphics[scale=0.35]{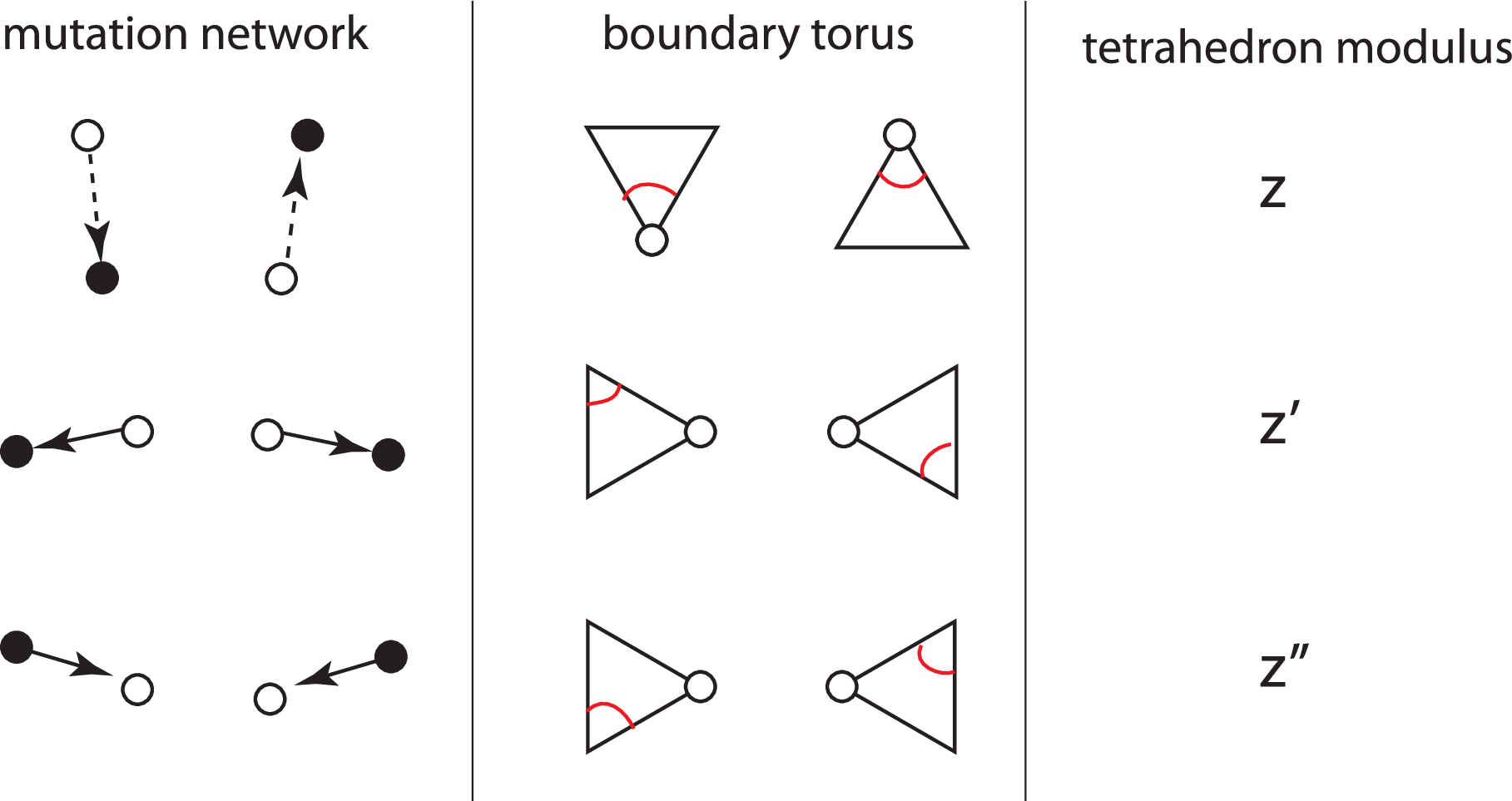}
  \caption{The mutation network specifies the figure of the boundary torus,
as well as the parametrization of the shape parameters of the
ideal tetrahedra. The three different types of edges correspond to 
three different parametrizations of a tetrahedron modulus.}
\label{fig.corr}
\end{figure}

The gluing equation has a rather concise expression in the mutation
network defined in section \ref{sec.network}.
The gluing equation can be written down for an internal edge 
of the triangulation, and
hence is associated with a white internal vertex of the network--- see the left of Figure
\ref{fig.gluetet}. 
The part of the mutation network around a white vertex
is in direct correspondence with the 
projection of shape of tetrahedra along the corresponding edge 
(Figure \ref{fig.gluetet}), or equivalently the boundary torus around
the edge.
Note that the mutation network also specifies the parametrization of
tetrahedra, i.e., whether we use $z, z'$ or $z''$
(Figure \ref{fig.corr}).

%%%%%%%%%%%%%%%%%%%%%%%%%%%%%%%%%%%%%%%%%%%%%
\subsection{Examples}\label{sec.example}
%%%%%%%%%%%%%%%%%%%%%%%%%%%%%%%%%%%%%%%%%%%%%

For concreteness and for comparison with earlier results,
we here work out two 3-manifold examples.
This will illustrate the generality and usefulness of our approach.

\subsubsection{Figure-Eight Knot Complement}

Let us first discuss one of the most famous hyperbolic knot complements, 
the figure-eight knot complement.
This can actually be realized as a mapping cylinder, and is discussed in
detail in Ref.\ \cite{Terashima:2011xe}. Here, we use the $4$-plat representation
of the knot.

\begin{figure}[htbp]
\centering\includegraphics[scale=0.33]{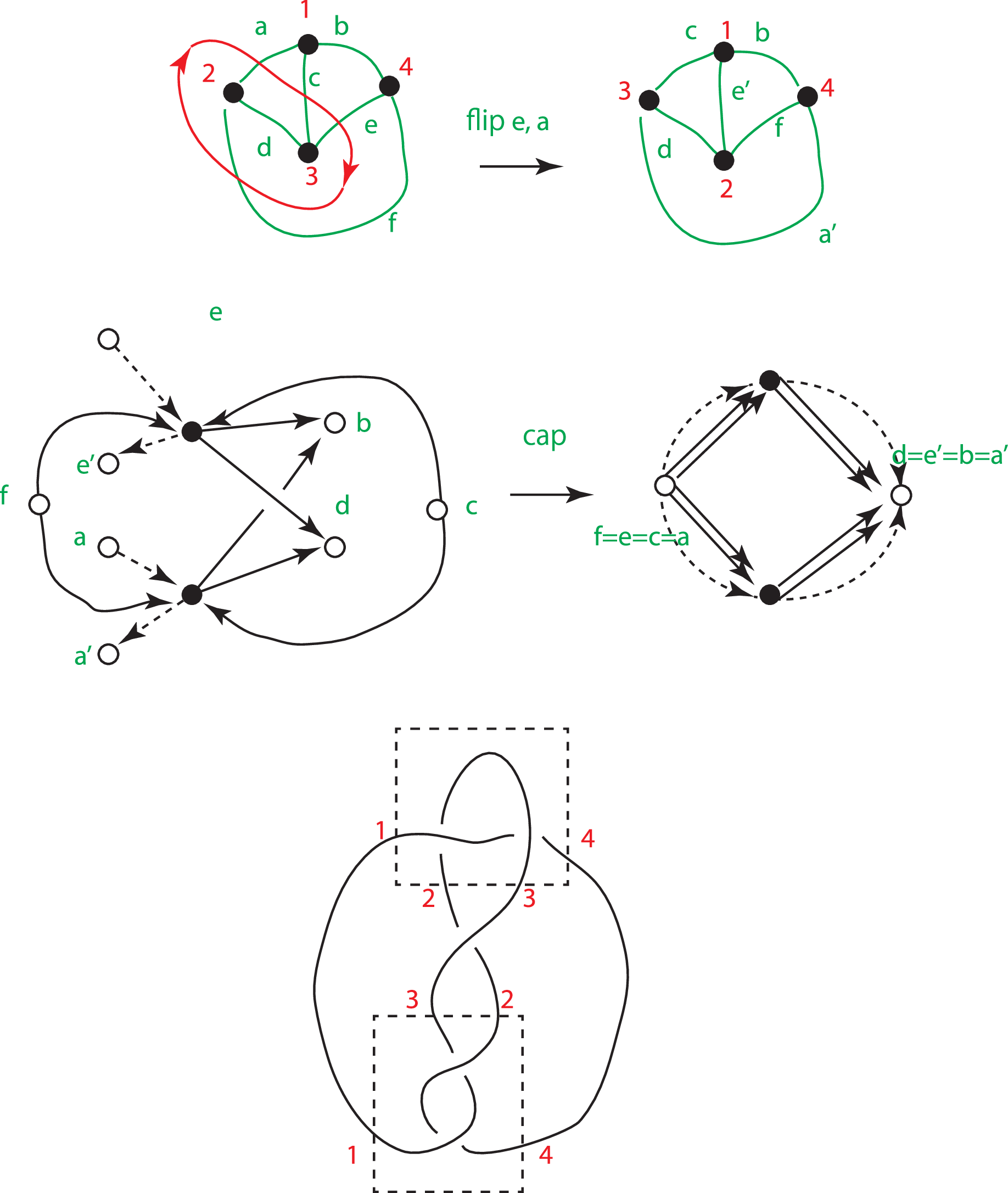}
\caption{The mutation sequence, the mutation network and the knot projection for the figure-eight knot complement $4_1$.
This gives the ideal triangulation in Figure \ref{fig.fig8trig}.}
\label{fig.fig8network}
\end{figure}

\begin{figure}[htpb]
\centering{\scalebox{0.9}{\includegraphics{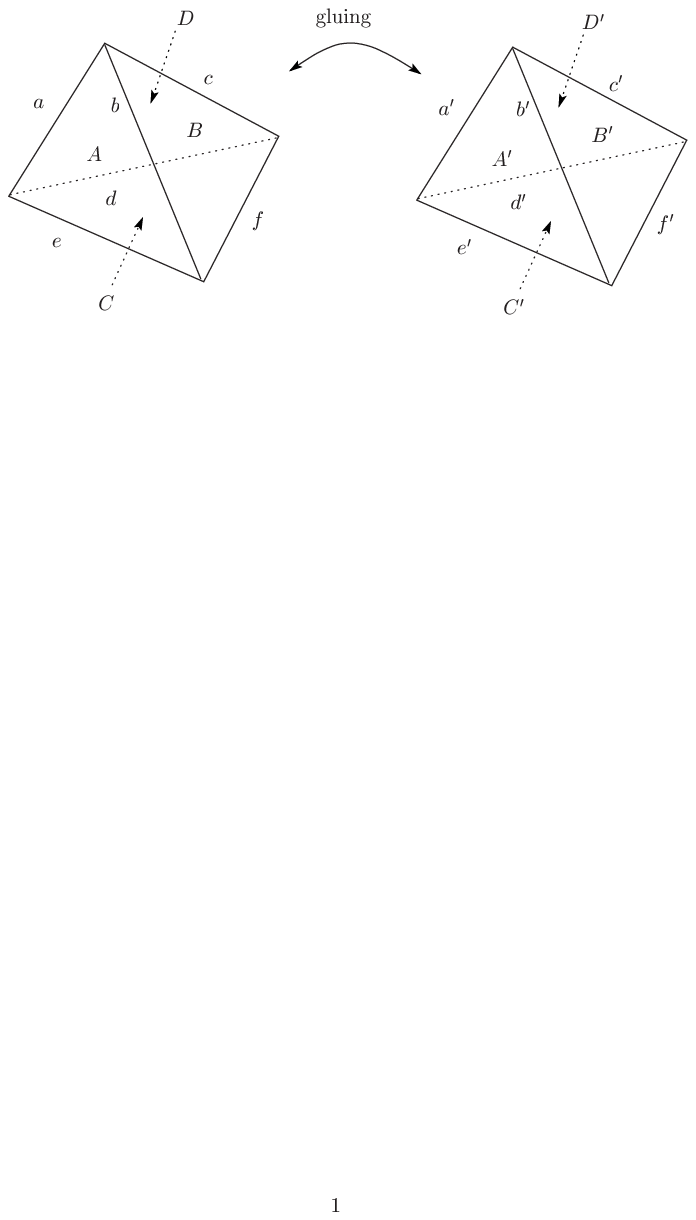}}}
\caption{An ideal triangulation of a figure-eight knot complement.}
%\vspace{0mm}
\label{fig.fig8trig}
\end{figure}

In the standard 4-plat representation of the figure-eight knot,
we need four generators of braid group $\scB_4$.
For practical computations, however, it is more efficient to 
incorporate some of the mapping class group actions into the 
choice of the caps (recall section \ref{sec.cap}).
We can then realize our knot by a single flip
with caps on both ends, 
leading to the mutation network in Figure \ref{fig.fig8network}.
Interestingly, this gives rise to the famous 
ideal triangulation of the figure-eight knot complement by the 
two ideal tetrahedra, found in 
 \cite{ThurstonLecture} (Figure \ref{fig.fig8trig}):
$$
A=A', \ B=B',\ C=C',\ D=D' \ ,
$$
with the identification of edges:
\begin{align*}
x:=a=d=f=a'=d'=f' \ , \quad
y:=b=c=e=b'=c'=e' \ .
\end{align*}

The semiclassical limit of the partition function \eqref{Wcontribution}
gives
\begin{align}
%\begin{split}
\scW(x,y)&=\Li(e^{2(-e-e'+b+d)})-\half \, 2(-e-e'+b+d)\, 2 (e+e'-c-f) \nonumber\\
&\ \ \ +\Li(e^{2(-a-a'+b+d)})-\half\, 2(-a-a'+b+d)\, 2 (a+a'-c-f) \nonumber \\
&=2(\Li(e^{2x-2y})+2(x-y)^2 )  \ ,
\end{align}
where $x, y$ are parameters associated with the two edges.
When we define $z:=e^{2x-2y}$,
the critical points are given by
$$
z^2-z+1=0 \ , \quad \textrm{i.e.}, \quad
z=\frac{-1\pm \sqrt{-3}}{2} \ .
$$
This corresponds the complete hyperbolic structure 
(and its complex conjugate) of 
the figure-eight knot complement.   
We can verify that the critical value of $\scW$ reproduces the hyperbolic
volume and the Chern--Simons invariant of the 3-manifold.

\begin{figure}[htbp]
\centering\includegraphics[scale=0.33]{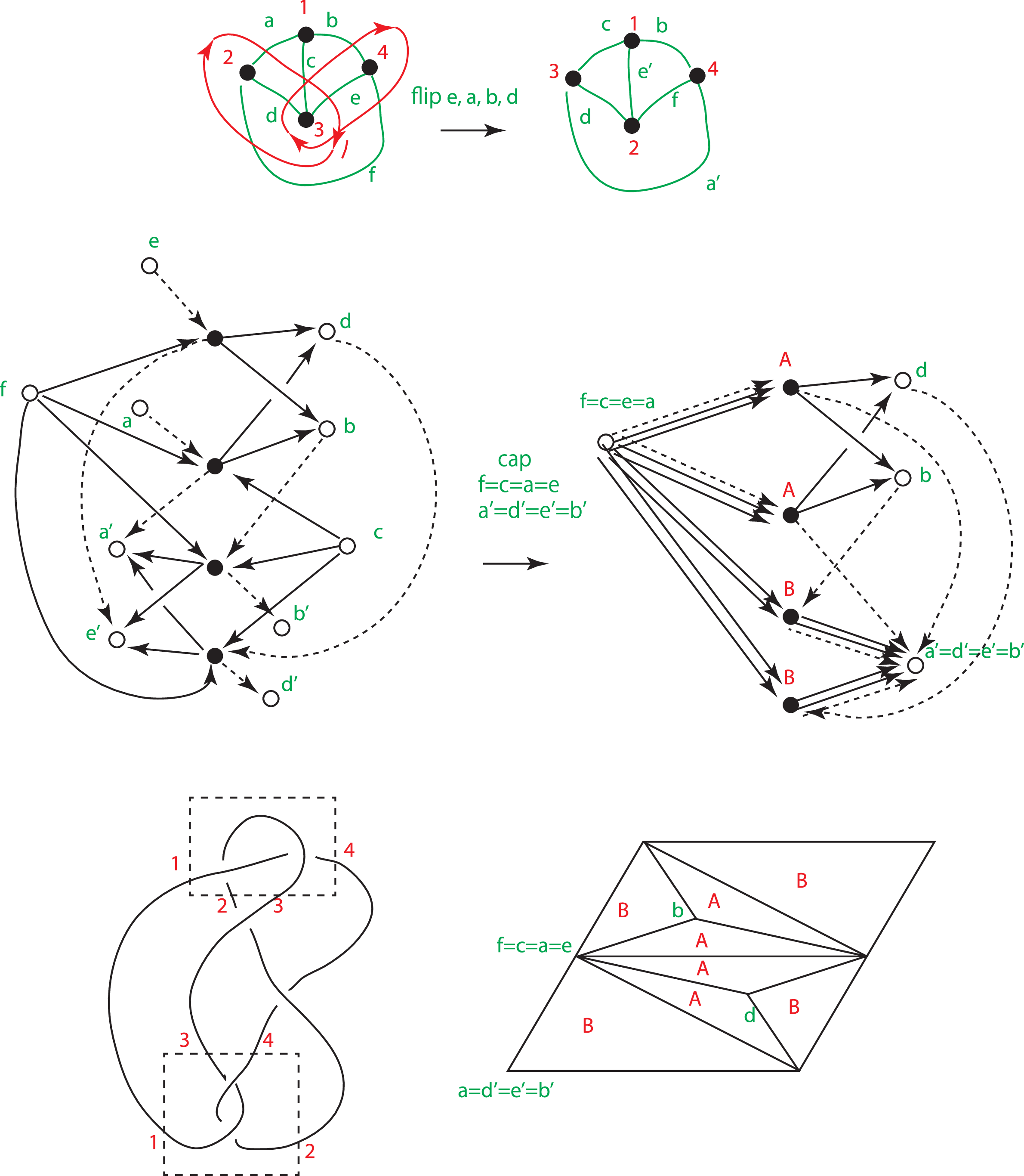}
\caption{The mutation sequence, the mutation network, knot projection 
and the boundary torus for the knot complement $5_2$. Compare this with \cite[Appendix]{FuterGueritaud}.}
\label{fig.52}
\end{figure}

By following similar methods, we can compute the partition function for
any link complements in $S^3$--- see Figure  \ref{fig.52} for another
example.
The general recipe for 
reading off a mutation sequence for a given Dehn twist is explained in
Appendix \ref{sec.Dehn}. This rule is rather useful
for practical computations.

%%%%%%%%%%%%%%%%%%%%%%%%%%%%%%%%%%
\subsubsection{Once-Punctured Torus Bundles Revisited}
%%%%%%%%%%%%%%%%%%%%%%%%%%%%%%%%%%%%

To illustrate the usefulness of our formalism,
let us work out the example of the once-punctured torus bundle.
This example has been worked out in detail in \cite{Terashima:2011xe}.
In particular it was found there that the quadratic piece of  
$\scW$ depends in a subtle way on the mutation sequence.
We find here that the rules proposed in this paper reproduce the findings
in \cite{Terashima:2011xe}.

\begin{figure}[htbp]
\centering{\includegraphics[scale=0.33]{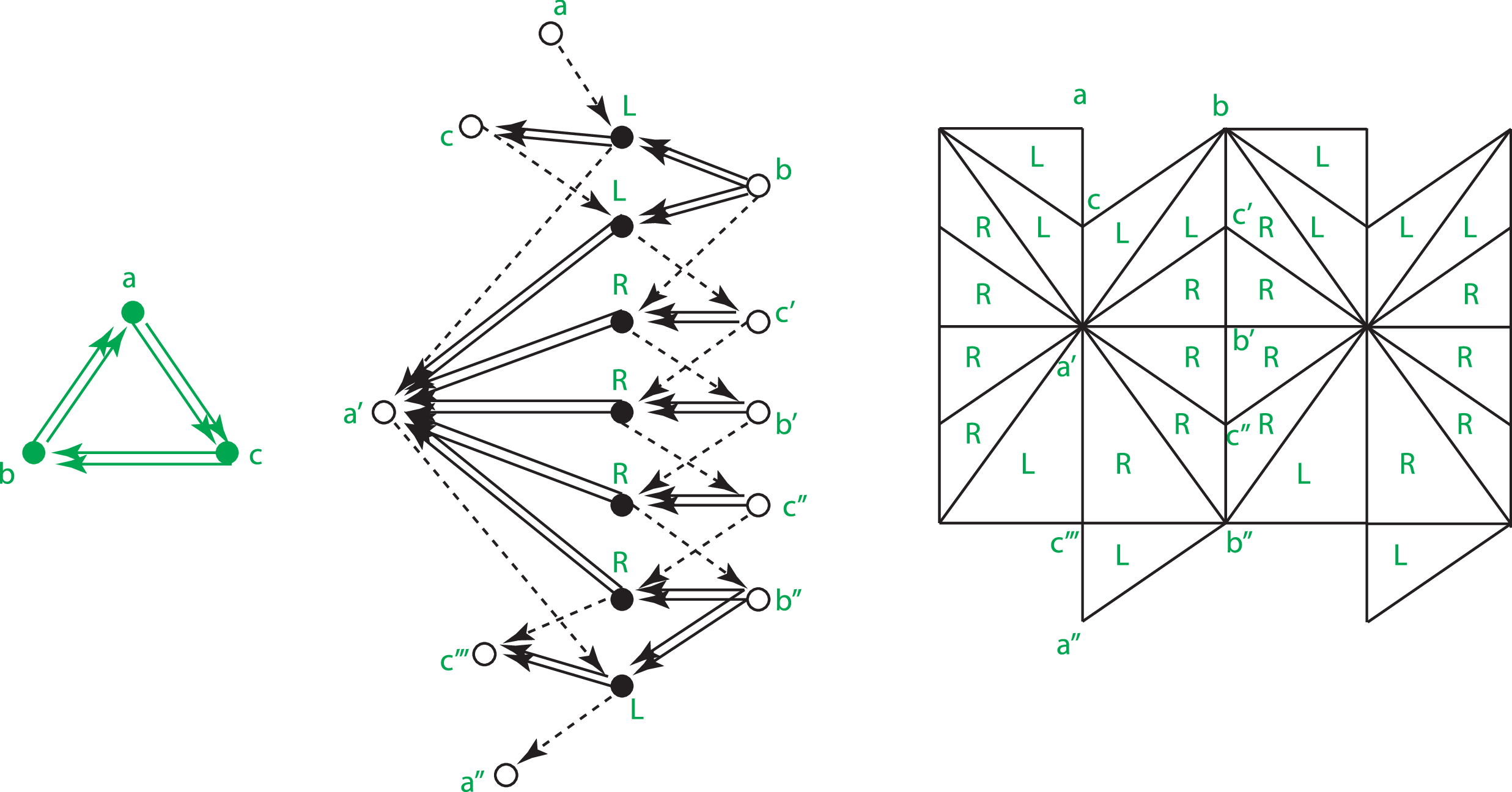}}
\caption{The quiver, the mutation network and the boundary torus
for a
once-punctured torus bundle with $\varphi=LLRRRRL$.
Compare the boundary torus with those in \cite{Gueritaud,Terashima:2011xe}.
}
\label{fig.oncetorus}
\end{figure}

Let us discuss the mapping cylinder for the once-punctured torus 
with $\varphi=LLRRRRL$, in the notation of \cite{Terashima:2011xe};
this is sufficient to discuss the 
general pattern.
The mutation network is given in Figure \ref{fig.oncetorus}.
We use the form of the semiclassical potential in \eqref{Wcontribution2},
\begin{equation}
\scW=\sum_{t=0}^{6} \left[ \Li_2(e^{Z(t)})-\frac{1}{2} (Z(t)-\i\pi)Z'(t) \right]\ ,
\end{equation}
with 
\begin{equation}
\begin{split}
&Z(0)-\i\pi= 2b-2c, \quad Z'(0)=-a-a'+2c,  \\
&Z(1)-\i\pi=2b-2a', \quad Z'(1)=-c-c'+2a',  \\
&Z(2)-\i\pi=2c'-2a', \quad Z'(2)=-b-b'+2a',  \\
&Z(3)-\i\pi=2b'-2a', \quad Z'(3)=-c'-c''+2a',  \\
&Z(4)-\i\pi=2c''-2a', \quad Z'(4)=-b''-b'''+2a',  \\
&Z(5)-\i\pi=2b''-2a', \quad Z'(5)=-c''-c'''+2a',  \\
&Z(6)-\i\pi=2b''-2c''', \quad Z'(6)=-a''-a'+2c'''
 \ ,
\end{split}
\end{equation}
where we have again used the simplified notation $w$ for the variable
$x_w$,
and for a mapping torus we identify $a''=a, b''=b, c'''=c$.
As have already seen, the saddle point of this potential $\scW$
gives the gluing equation of the hyperbolic 3-manifold.

Let us define $w_t:=Z(t-1)-\i \pi$ for $t=1, \ldots, 7$,
and re-express $\scW$ in terms of the $w_t$s.
We then find
\begin{equation}
\begin{split}
\scW(w)=&\Li_2(-e^{w_1})+\frac{1}{4}w_1(-w_7-w_2+2w_1) \\
&+\Li_2(-e^{w_2})+\frac{1}{4}w_2(-w_1+w_2+w_3) \\
&+\Li_2(-e^{w_3})+\frac{1}{4}w_3(w_2+w_4) \\
&+\Li_2(-e^{w_4})+\frac{1}{4}w_4(w_3+w_5) \\
&+\Li_2(-e^{w_5})+\frac{1}{4}w_5(w_4+w_6) \\
&+\Li_2(-e^{w_6})+\frac{1}{4}w_6(w_5-w_7+w_6) \\
&+\Li_2(-e^{w_7})+\frac{1}{4}w_7(-w_6-w_1+2w_7)  \ .
\label{eq135}
\end{split}
\end{equation}
Note that associated with the $\Li_2(-e^{w_t})$, there is a quadratic
expression with respect to $w_{t-1}, w_{t}, w_{t+1}$.
This is determined by the choice of either $L$ or $R$
for the neighboring $(t-1), t, (t+1)$-th flips, which for
$t=1, \ldots, 7$ are given by
\begin{equation*}
1: LLL \ , \quad 2: LLR \ , \quad 3: LRR \ ,\quad
4: RRR \ , \quad 5: RRR \ ,\quad 6: RRL \ ,\quad
7:RLL \ . 
\end{equation*}
We can compare this with the results of \cite{Terashima:2011xe},
and find complete agreement.
For example, for the first line of \eqref{eq135}
we have $LLL$, and this coincides with the case 1 of
\cite[section 3.4]{Terashima:2011xe}.

As this example illustrates, 
while it is possible to write the final expression only 
in terms of the cluster $y$-variable,
the most systematic expression requires the use of the 
cluster $x$-variables.

%%%%%%%%%%%%%%%%%%%%%%%%%%%%%%%%%%%%%%%%%%%%%
\section{Summary and Outlook}\label{sec.conclusion}
%%%%%%%%%%%%%%%%%%%%%%%%%%%%%%%%%%%%%%%%%%%%%

In this paper, we have proposed a new conjecture that a class of 
3d $\scN=2$ theories is naturally and systematically associated with a sequence of quiver
mutations. The data of the quiver mutation is encoded in a bipartite
graph, the mutation network, from which we have computed the associated
cluster partition function %(Figure \ref{fig.flow}) 
and identified the
corresponding
3d $\scN=2$ theory. The rules are summarized in Table \ref{table.summary}. 
This is a rather general procedure, and includes in particular theories
associated with 3-manifolds.

\begin{table}[htbp]
\caption{Dictionary between mutation network, 3-manifold and 3d $\scN=2$
 theory.}
\label{table.summary}
{\renewcommand\arraystretch{1.4}
\begin{center}
\begin{tabular}{c|c|c}
 mutation network & 3-manifold & 3d $\scN=2$ theory  \\ 
\hline
\hline
a white vertex & an edge of tetrahedron  & a $U(1)$-symmetry \\
\hline
an intermediate white vertex     & an internal edge & a gauge $U(1)$-symmetry  \\
\hline
an initial/final white vertex    & a boundary edge & a global $U(1)$-symmetry \\
\hline
a parameter associated &  a parameter on & a vector multiplet scalar \\ 
with a white vertex &  an edge of a tetrahedron &  \\
\hline
a black vertex   & an ideal tetrahedron & an $\scN=2$ chiral multiplet \\
\hline
an edge connecting & an edge belonging to & $U(1)$ charges  \\
black and white vertices & an ideal tetrahedron & of a chiral multiplet \\
\hline
black vertices& ideal tetrahedra & a superpotential term \\
connected to a white vertex & glued around an edge & \\
%\hline
\end{tabular}
\end{center}
}
\end{table}

We leave the detailed field theory analysis of our theories
for future work.
For example, it would be interesting to discuss 
the mutations of the $(A_k, A_n)$ quivers discussed in
\cite{Cecotti:2010fi,Xie:2012mr,Heckman:2012jh,Franco:2012mm}.

In this paper we have
identified our 3d $\scN=2$ theories
based on the relations \eqref{ZZ}, \eqref{ZTQ}
and the $S^3_b$ partition function.
For the case with 3-manifolds, 
it is believed that our 3d 
$\scN=2$ theories
actually
arise from compactification of 6d $(2,0)$
theory on a 3-manifold, and also from the boundary conditions of
4d $\scN=2$ Abelian theories.
The latter in particular gives a direct 
physical method for identifying the 3d $\scN=2$ theories,
which are expected to have the same $S^3_b$ partition function as our 3d Abelian theories.
This program has been carried out for 1/2 BPS boundary conditions in
4d $\scN=4$ theory \cite{Gaiotto:2008sa,Gaiotto:2008ak}, and 
more results in this direction will appear in
the upcoming work \cite{HPY}.

There are also more mathematical questions to ask --- our partition
function defines a knot invariant, and it would be desirable to 
define the invariant more rigorously.
In fact, the discussion in 
section \ref{sec.3mfd} uses braid groups and plat representations of knots,
which is often used in the study of Jones polynomials and their
generalizations (cf.\ \cite{ReshetikhinTuraev}).

For the case without a 3-manifold description,
we have different questions to ask. Why does the relation \eqref{ZTQ}
hold? For precisely which class of 3d gauge theories does the relation
hold? Do we have a string theory realization of our theories? 
As noted previously, our discussion includes 
the case where our quiver is identified with the BPS quivers
of 4d $\scN=2$ theories, and for these examples 
it is expected that our 3d $\scN=2$ theories
are the 1/2 BPS boundary theories of the 4d $\scN=2$ theories.
However, our quivers in this paper can be arbitrary quivers
and they will not necessarily be the BPS quivers.

The fact 
that \eqref{ZTQ} holds for a rather rich class of 3d
$\scN=2$ theories is a strong indication that there exists a 
rather rich structure in the ``space of 3d $\scN=2$ theories''
beyond the realm of 3-manifold theories, and 
what we know right now is probably only a tip of the iceberg
of a much richer structure.
Indeed, the cluster algebra in our paper, and their interpretation as the algebra of loop operators, suggest
the general philosophy that \emph{the IR fixed points of 3d $\scN=2$ theories can be characterized by the algebra of 
1/2-BPS loop operators}.

One important clue for this ambitious program should be the mathematical
structures discussed in this paper, such as cluster algebras
and hyperbolic 3-manifolds.
They appear in diverse areas of physics and mathematics,
including
wall-crossing phenomena of 4d $\scN=2$ theories \cite{Cecotti:2011rv,Alim:2011kw,Xie:2012gd},
dimer integrable models \cite{Goncharov:2011hp,Franco:2011sz},
on-shell scattering amplitudes \cite{ArkaniHamed:2012nw},
and superpotential conformal indices of 4d $\scN=1$ theories and their
dimensional reductions, as well as associated integrable spin lattices \cite{Terashima:2012cx,Yamazaki:2012cp}.

%%%%%%%%%%%%%%%%%%%%%%%%%%%%%%%%%%%%%%%%%%%%%
\section*{Acknowledgments}
We would like to thank N.\ Drukker, K.\ Hosomichi, Y.\ Imamura, 
R.\ Inoue, K.\ Ito, T.\ Okuda, S.\ Terashima, D.\ Yokoyama
and in particular K.\ Nagao for discussion.
M.~Y.\ would like to thank in particular D.\ Xie 
for related discussion.
M.~Y.\ would also like to thank the Aspen Center for Physics (NSF Grant No.\ 1066293), 
the Kavli Institute for Theoretical Physics, the Newton Institute (Cambridge University),
and the Simons Center for Geometry for Physics (Simons Summer Workshops in
Mathematics and Physics 2011 and 2012) and 
Yukawa Institute for Theoretical Physics (YKIS 2012)
for hospitality during various stages of this project.
Part of the contents of this paper have been presented by M.~Y.\ during 
seminars and conferences in a number of universities and research
institutes, including the SPOCK meeting (University of Cincinnati), Nov.\ 2011; 
Brown University, Dec.\ 2011; 
IPMU (University of Tokyo), Jan.\ 2012; University of Cambridge, Feb.\
2012, and in particular RIMS, Kyoto University, Jan.\ 2013;
``Exact Results in SUSY Gauge Theories and Integrable Systems,'' Rikkyo
University, Jan.\ 2013.
We thank the audience for invaluable feedback and advice.

%%%%%%%%%%%%%%%%%%%%%%%%%%%%%%%%%%%%%%%%%%%%%%%%%%%%%%%%%%%%%%%%%%%%%%
\appendix 

%%%%%%%%%%%%%%%%%%%%%%%%%%%%%%%%%%%%%%%%%%%%%%%%%%%%%%%%%%%%%%%%%%%%%%%%%%%%%
\section{Quantum Dilogarithms} \label{sec.dilog}

In this appendix we collect formulas for the so-called non-compact quantum
dilogarithm function (simply called quantum dilogarithm function
in the main text) $s_b(z)$ and $e_b(z)$ \cite{FaddeevVolkovAbelian,FaddeevKashaevQuantum,Faddeev95}. 

We define the function $s_b(z)$ by
\begin{align}
s_b(z)=\exp\left[ \frac{1}{i} \int_0^{\infty} \frac{dw}{w} 
\left( \frac{\sin 2zw}{2\sinh (bw) \sinh (w/b)}-\frac{z}{w} \right)
 \right]
\ ,
\label{sbdef}
\end{align}
and $e_b(z)$ by
\begin{align}
e_b(z)=\exp\left(\frac{1}{4} \int_{-\infty+i0}^{\infty+i0}  
\frac{dw}{w} \frac{e^{-i 2zw} }{\sinh(wb) \sinh(w/b)}\right), 
\label{ebdef}
\end{align}
where the integration contour is chosen above the pole $w=0$. 
In both these expressions we require $|\mathrm{Im}\, z| <|\mathrm{Im}\,
c_b|$ for convergence at infinity. The
two functions are related by
\begin{align}
e_b(z)=e^{\frac{\pi i z^2}{2}} e^{- \frac{i\pi (2-Q^2)}{24}}s_b(z) \ ,
\label{ebsb}
\end{align}
where $\scQ:=b+b^{-1}$.

It immediately follows from the definition that
\begin{align}
e_b(z)=e_{b^{-1}}(z)\ , \quad 
s_b(z)=s_{b^{-1}}(z) \ , \quad
s_b(z) s_b(-z)=1 \ .
\label{sbinverse}
\end{align}

In the classical limit $b\to 0$, we have
\begin{align}
e_b(z)\to \exp \left(\frac{1}{2 \pi i b^2} \textrm{Li}_2(-e^{2\pi b z})
\right),
\label{ebclassical}
\end{align}
where $\textrm{Li}_2(z)$ denotes classical dilogarithm function of Euler,
defined by
\begin{align}
\textrm{Li}_2(z)=\sum_{n=1}^{\infty} \frac{z^n}{n^2}=-\int_0^z
\frac{\log(1-t)}{t}dt \ . 
\label{Li2}
\end{align}

The Fourier transform of $e_b(z)^{\pm 1}$ basically comes back to itself \cite{Faddeev:2000if}:
\begin{align}
\begin{split}
&\int dx\, e_b(x)\, e^{2\pi i  w x}=e^{-i\pi w^2+\frac{i\pi}{12} (1+Q^2)}
\, e_b\left(w+i \frac{Q}{2}\right)
\ , \\
& \int dx\, e_b(x)^{-1}\, e^{2\pi i  w x}=e^{i\pi w^2- \frac{i\pi }{12}(1+Q^2)}
\, e_b\left(-w-i \frac{Q}{2}\right)^{-1}
\ .
\end{split}
\label{ebFourier}
\end{align}

%%%%%%%%%%%%%%%%%%%%%%%%%%%%%%%%%%%%%%%%%%%%%
\section{Hyperbolic Geometry in a Nutshell}\label{sec.brief}
%%%%%%%%%%%%%%%%%%%%%%%%%%%%%%%%%%%%%%%%%%%%%%%

\begin{figure}[htbp]
\centering{\includegraphics[scale=0.28]{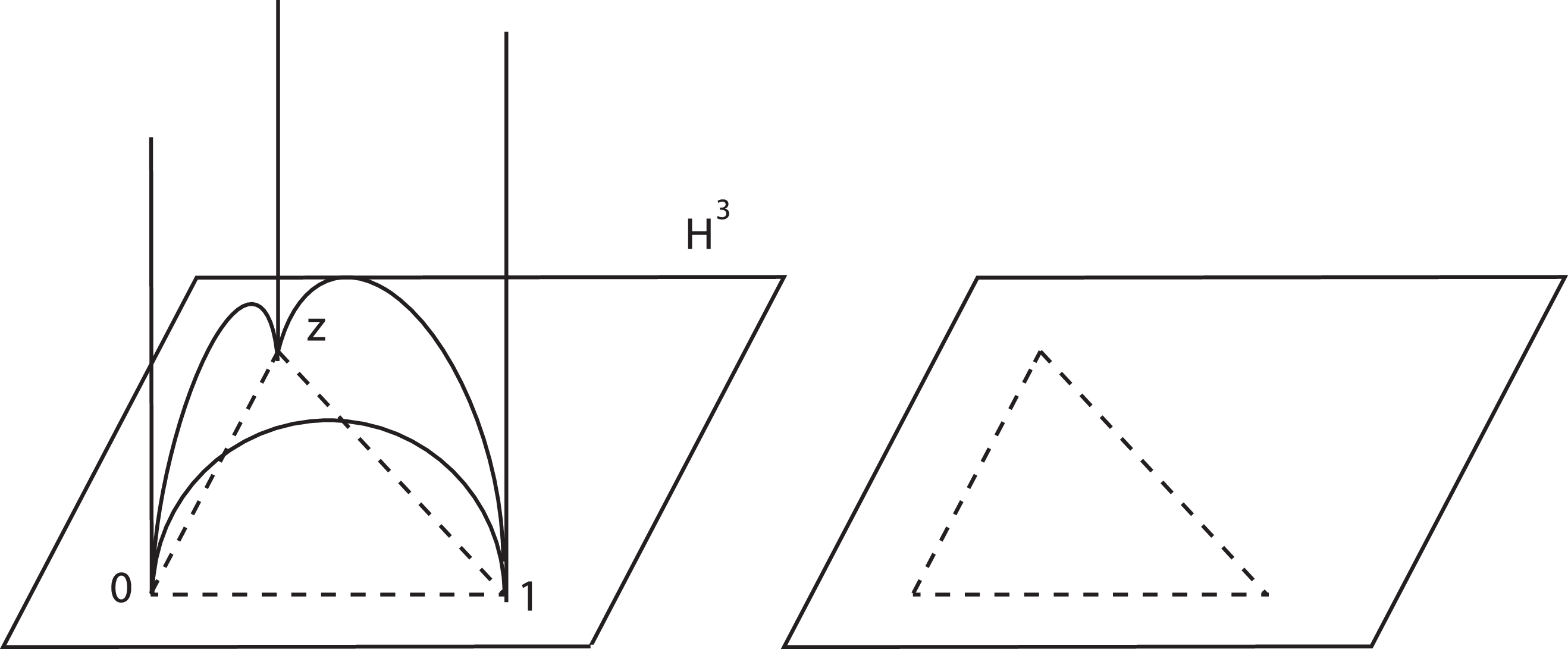}}
\caption{An ideal tetrahedron in $\bH^3$ 
has all four vertices on
 the boundary of $\bH^3$,
which we can take to be 
$\{0,1,z,\infty\} \in
 \bC\cup \{\infty\}$.}
\label{idealtetrahedron}
\end{figure}

In this appendix we 
briefly recall the minimal ingredients of
classical 3d hyperbolic geometry 
(see, e.g., \cite{ThurstonLecture,ThurstonReview}), 
for readers unfamiliar with the subject.
Let $\bH^3$ be the 3d hyperbolic space, namely the upper half plane
\begin{align}
\bR^3_{> 0}=\{(x_1, x_2, y) |\, x_1, x_2\in \bR, y>0 \} \, ,
\end{align}
with the
metric
\begin{align}
ds^2=\frac{(dx_1)^2+(dx_2)^2+dy^2}{y^2} \, .
\end{align}  
An ideal tetrahedron is a tetrahedron all four of whose vertices are on the
boundary of $\bH^3$ (see Figure \ref{idealtetrahedron}). By a suitable
isometry of $\bH^3$ we can take the vertices to be
at positions $0,1,z$ and infinity. This complex parameter $z$ is called the
modulus (shape parameter) of the tetrahedron.

A tetrahedron has six edges and, correspondingly, six face angles. For an ideal
tetrahedron with modulus
$z$, the two face angles on the opposite side of the tetrahedron are
the same. These are given as the arguments of three complex parameters
\begin{equation}
z\ , \quad z':=\frac{1}{1-z} \ , \quad z'':=
1-z^{-1}\, ,
\label{ztriple}
\end{equation}
satisfying $z z' z''=-1$.
These three distinct parametrizations of the tetrahedron modulus will 
play an important later, when we discuss our rules. 

When we glue the tetrahedra to construct 3-manifolds, we need to ensure
that the angles around an edge sum up to $2\pi$. Since the angles
of tetrahedra could be described either by $z, z'$ or $z''$ depending on
the parametrization, we have
\begin{align}
\prod_{m:\, {\rm type\ 1}} z^{(m)}
\prod_{m:\, {\rm type\ 2}} z'^{(m)}
\prod_{m:\, {\rm type\ 3}} z''^{(m)}
=1 \ ,
\label{structure}
\end{align}
where we classified a tetrahedron $m$ depending on whether the angle around the
edge is parametrized by $z^{(m)}, z'^{(m)}$ or $z''^{(m)}$.
We call these equations \emph{gluing equations}.\footnote{They are also called structure equations.}
In the boundary torus this simply represents the the condition that
the product of $z$ around a vertex is trivial. 
We refer to this equation in section \ref{sec.structure}.

%%%%%%%%%%%%%%%%%%%%%%%%%%%%%%%%%%%%%%%%%%%%%
\section{Dehn Twists as Flips}\label{sec.Dehn}
%%%%%%%%%%%%%%%%%%%%%%%%%%%%%%%%%%%%%%%%%%%%%

As explained in the main text, an element of the mapping class group
could be represented as a sequence of flips on the 2d triangulation. 
In this appendix we identify this flip
sequence systematically. 
This result will be important for practical computations.

\begin{figure}[htbp]
\centering{\includegraphics[scale=0.45]{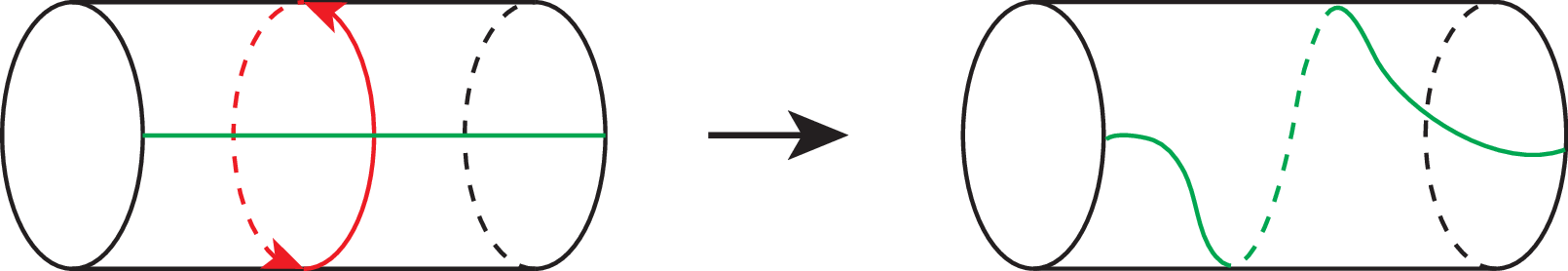}}
\caption{
A Dehn twist along a non-contractible cycle (colored red).
}
\label{fig.Dehn}
\end{figure}

The mapping class group is generated by 
the Dehn twist $D_{\gamma}$ along non-contractible cycles 
(Figure \ref{fig.Dehn}).
Moreover explicit generators and relations for the
mapping class group of $\Sigma_{g,h}$
are known in the literature (see e.g.\ \cite{Birman,Gervais}).
This means that all we need to do is to identify an explicit sequence of
flips corresponding to a single Dehn twist.

\begin{figure}[htbp]
\centering{\includegraphics[scale=0.33]{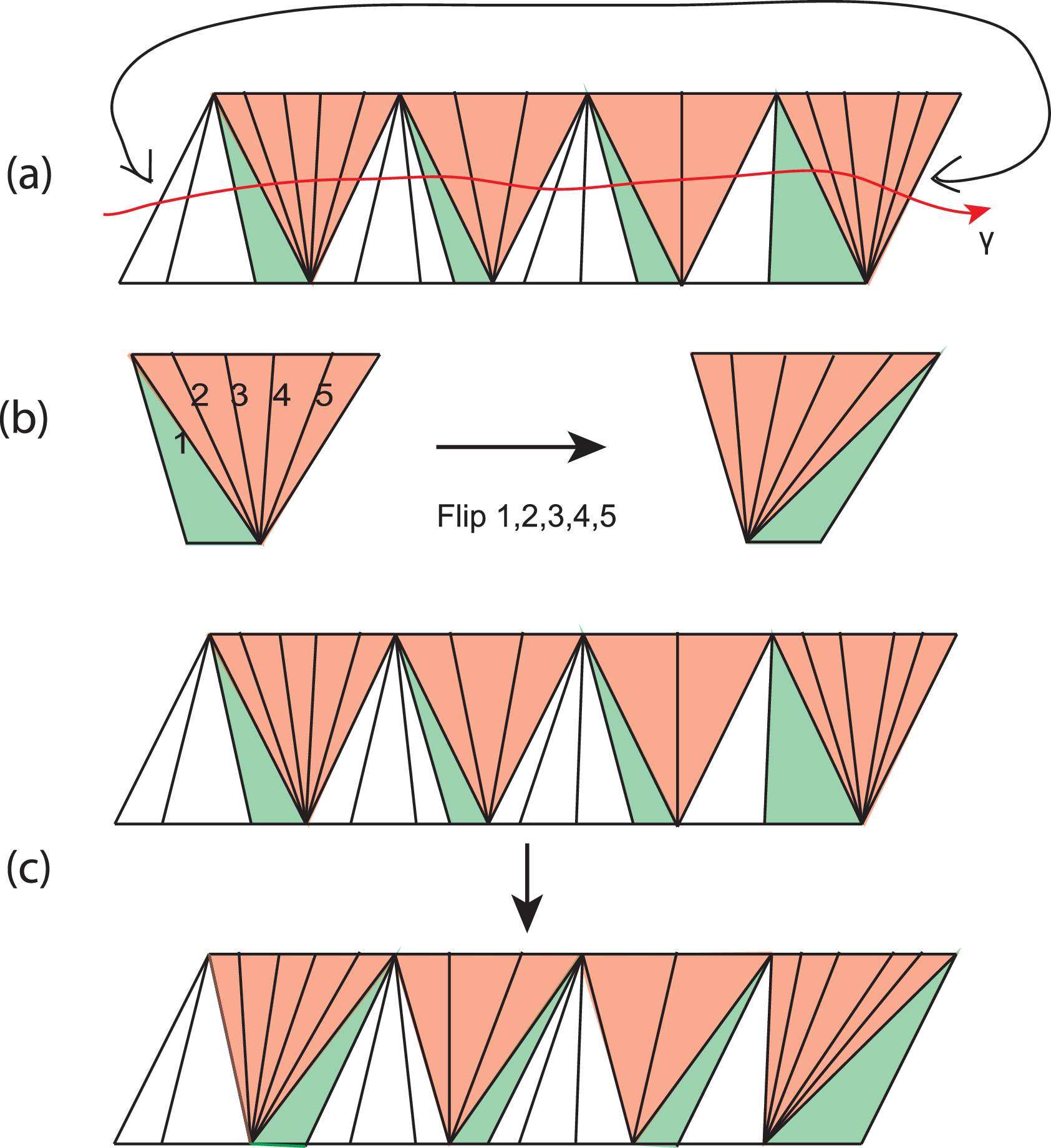}}
\caption{A flip sequence for a Dehn twist along a cycle $\gamma$.
(a) The cycle $\gamma$ intersects many triangles along the way, 
some of which have an extra edge on the left and some on the right.
We have highlighted all the triangles on the left side (colored red)
and some on the right (colored green).
(b) For each red/green block, we perform a sequence of flips, as here.
(c) By repeating (b), and effectively shifting the position of one
 triangle one by one, we obtain a Dehn twist on the triangulation.
}
\label{fig.annulusDehn}
\end{figure}

Suppose that we perform a Dehn twist along a
non-contractible cycle $\gamma$, which intersects several
triangles as in Figure \ref{fig.annulusDehn} (a).
We can then verify that the flips shown in Figure \ref{fig.annulusDehn}
(b) and (c) realize the Dehn twist. Note also we need to exchange the labels
appropriately after the flips.
This is a rather general rule, which applies to any triangulation
and to surfaces of any genus.

The situation is especially simple in the case of the $1$-punctured torus, studied in detail
in \cite{Terashima:2011qi,Terashima:2011xe} 
--- the Dehn twist along the $\alpha, \beta$-cycles 
both correspond to a
single flip.

\begin{figure}[htbp]
\centering{\includegraphics[scale=0.33]{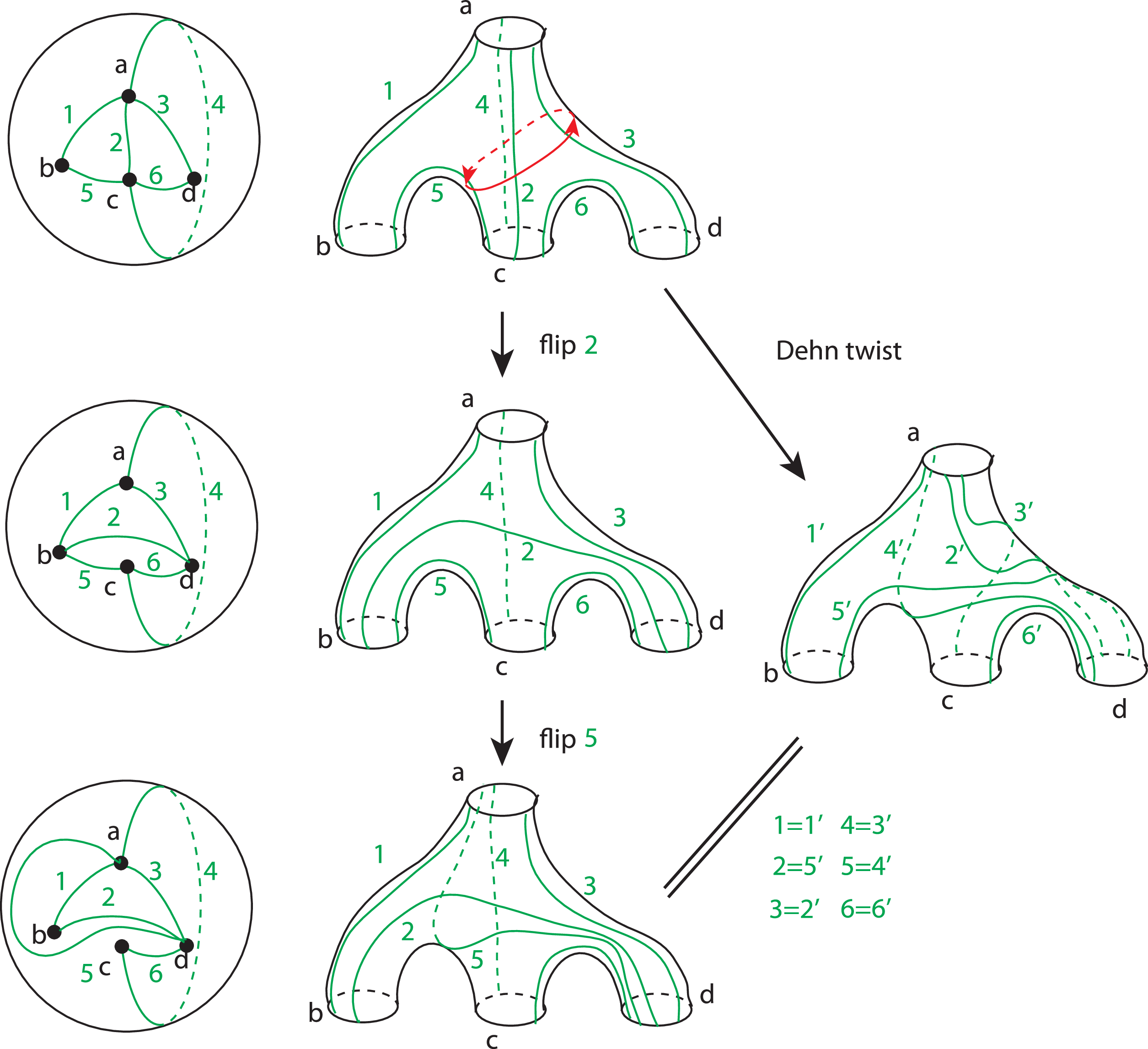}}
\caption{A Dehn twist along the red cycle could be traded for two flips
 (and re-labeling).}
\label{fig.4sphere}
\end{figure}

\begin{figure}[htbp]
\centering{\includegraphics[scale=0.27]{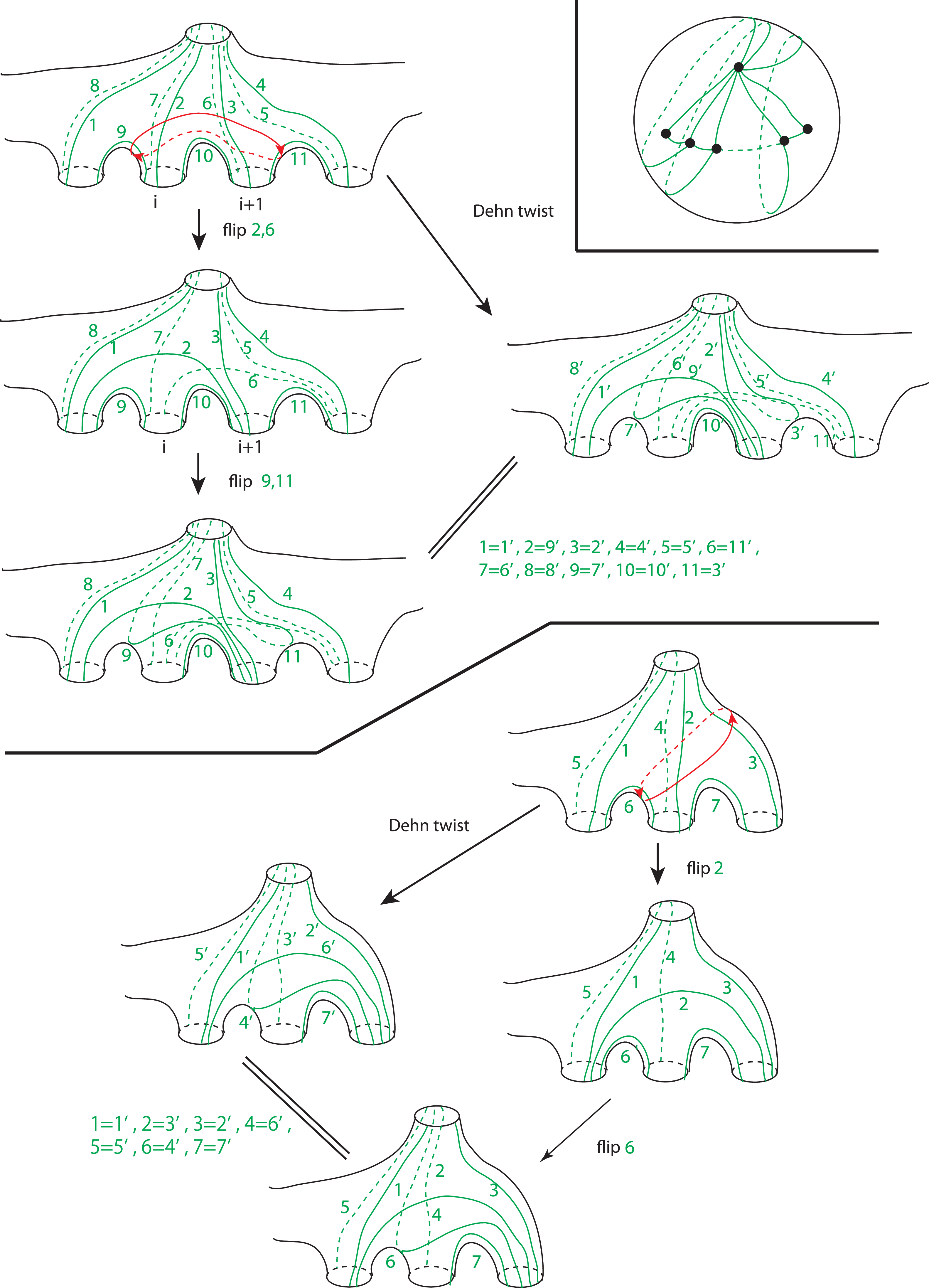}}
\caption{The Dehn twist ($s_i$) on the $n$-punctured
 sphere could be traded for two or four flips
(and re-labeling).
}
\label{fig.nsphere}
\end{figure}

For the $n$-punctured sphere,
consider a cycle
encircling the $i$-th and $(i+1)$-th 
punctures, and the corresponding Dehn twist $s_i$.
These Dehn twists generate the braid group.
We can verify that this Dehn twist $s_i$,
in the triangulations in Figures \ref{fig.4sphere} and \ref{fig.nsphere},
are given by either $2$ or $4$ flips. 

\clearpage

%%%%%%%%%%%%%%%%%%%%%%%%%%%%%%%%%%%%%%%%%%%%%%%%%%%%%%%%%
%%%%%%%%%%%%%%%%%%%%%%%%%%%%%%%%%%%%%%%%%%%%%%%%%%%%%%%%%

\bibliographystyle{JHEP}
\bibliography{../Heegaard/AGTknot}

\end{document}